\documentclass[12pt] {article} 
\usepackage {amsmath, amssymb, graphicx, multirow, rotating}
\usepackage{hyperref}
 \usepackage{multicol}

\numberwithin{equation}{section}

\newcommand{\ack}{\par\nobreak\medskip\nobreak
\begin{center}{\bf Acknowledgments}\end{center}
\par\nobreak\nobreak}

\makeatletter
\def\@seccntformat#1{\csname the#1\endcsname.\quad} 

\renewcommand\section{\@startsection
  {section}{1}{0mm}%name, level, indent
  {-\baselineskip}%             beforeskip
  {0.5\baselineskip}%            afterskip
  {\normalfont\normalsize\bf}}%{\hspace{-.45cm}.\; #1}}% style

\renewcommand\subsection{\@startsection
  {subsection}{2}{0mm}%name, level, indent
  {-\baselineskip}%             beforeskip
  {0.5\baselineskip}%            afterskip
  {\normalfont\normalsize\it}}%{\hspace{-.45cm}.\; #1}}% style
\makeatother

\newcommand{\drawsquare}[2]{\hbox{%
\rule{#2pt}{#1pt}\hskip-#2pt%  left vertical
\rule{#1pt}{#2pt}\hskip-#1pt%  lower horizontal
\rule[#1pt]{#1pt}{#2pt}}\rule[#1pt]{#2pt}{#2pt}\hskip-#2pt%  upper  horizontal
\rule{#2pt}{#1pt}}% right vertical

\newcommand{\fund}{\drawsquare{7}{0.6}}%  fundamental
\newcommand{\asymm}{\drawsquare{7}{0.6}\hskip-7.6pt%
     \raisebox{7pt}{\drawsquare{7}{0.6}}}%  antisymmetric second rank

\newcommand{\unit} [1] {\; \mathrm {#1}}

\newcommand{\cO}{\mathcal{O}}
\newcommand{\Z}{{\mathbb Z}}
\newcommand{\W}{\mathcal{W}}
\newcommand{\Wdark}{\mathcal{W}_{d}}
\newcommand{\tr}{\mathrm{Tr}}
\newcommand{\mgut}{M_{\rm GUT}}
\newcommand{\Gdark}{G_{d}}
\newcommand{\kahler}{{K\"ahler} }
\newcommand{\alphaEM}{\alpha_{\rm EM}}
\newcommand{\mdark}{m_{\gamma_d}}
\newcommand{\Mp}{M_{\rm Pl}}
\newcommand{\mdm}{m_\chi}
\newcommand{\mgaugino}{ m_{\tilde \gamma_d}}
\newcommand{\Udark}{U(1)_{d}}
\newcommand{\dgauge}{\gamma_d}
\newcommand{\dgaugino}{\tilde \gamma_d}
\newcommand{\gdark}{g_{d}}
\newcommand{\alphadark}{\alpha_{d}}
\newcommand{\SUtwodark}{SU(2)_d}
\newcommand{\Utev}{U(1)_\chi}
\newcommand{\SUfive}{SU(5)_{\rm SM}}

\begin{document}
\begin{titlepage}
\begin{flushright}
%{\tt hep-ph/0712.xxxx
%}
\end{flushright}

\vskip.5cm
\begin{center}
{\LARGE Decaying into the Hidden Sector}
\vspace{.2cm}
 
\vskip.2cm
\end{center}
\vskip0.2cm

\begin{center}
{\bf Joshua T. Ruderman${}^a$ and Tomer Volansky${}^b$}
\end{center}
\vskip 8pt

\begin{center}
{\it ${}^a$  Department of Physics, Princeton University, Princeton,
  NJ 08544} \\
{\it ${}^b$  Institute for Advanced Study, Princeton, NJ 08540} \\
\vspace*{0.3cm}
\end{center}

\vglue 0.3truecm

\begin{abstract}
  \vskip 3pt \noindent The existence of light hidden sectors is an
  exciting possibility that may be tested in the near future.  If DM is allowed to decay into such a hidden sector through GUT suppressed operators, it can accommodate the recent cosmic ray observations without over-producing antiprotons or interfering with the attractive features of the thermal WIMP.   Models of this kind are simple to construct, generic and evade all astrophysical bounds.  We provide
  tools for constructing such models and present several distinct examples.  
  The light hidden spectrum and DM
  couplings can be probed
  in the near future,
  by measuring astrophysical photon and neutrino fluxes.  These indirect signatures are complimentary to the direct production 
signals, such as lepton jets, predicted by these models.
\end{abstract}

\end{titlepage}

\tableofcontents

\newpage

\section{Introduction}
\label{sec:introduction}

The existence of a low energy hidden sector, weakly coupled to the
Standard Model (SM), is an exciting possibility that will be tested by
upcoming experiments.  Hidden sector particles can be produced in high
energy colliders, as stressed in the context of `Hidden Valley'
models~\cite{Strassler:2006im} and models where gauge kinetic mixing
results in `lepton
jets'~\cite{ArkaniHamed:2008qp,Baumgart:2009tn}. Such hidden sectors
can also be probed with low energy $e^+ e^-$ colliders and fixed
target experiments~\cite{Pospelov:2008zw}.  Here, we point out that
the existence of a low energy hidden sector, together with weakly
interacting DM (WIMP) and gauge coupling unification, implies the
generic possibility that DM may decay directly into the hidden sector
through operators suppressed by the GUT scale.  These decays, followed by
decays into SM particles through kinetic mixing, provide the
intriguing possibility of using astrophysical observations to study the
hidden sector spectrum, complementing direct production experiments.
This decaying DM framework provides a simple and natural explanation for the
recent cosmic ray (CR) anomalies~\cite{Adriani:2008zr}, while avoiding
the tensions and pitfalls of many previously proposed models.

A DM explanation of the electronic CR excess requires a DM mass
greater than a TeV~\cite{Meade:2009iu,Bergstrom:2009fa}, and
predominantly leptonic production~\cite{Barger:2008su}.
Consequently, the vanilla
MSSM WIMP scenario is disfavored, and many new models have been
proposed, bifurcating  into annihilating models~\cite{ArkaniHamed:2008qn,
  Annihilate} and decaying models~\cite{Eichler:1989br, Nardi:2008ix,
  Arvanitaki:2008hq,Chen:2009iua, Buchmuller:2007ui}.  Annihilating
models are difficult to reconcile with the FERMI and HESS CR data,
because the softening of the spectrum above a few TeV requires an
annihilation cross-section $\cO(1000)$ times larger than that of the
standard thermal WIMP~\cite{Meade:2009iu,Bergstrom:2009fa}.  Such a
large cross-section is in tension with constraints from photon and
neutrino measurements from the Galactic Center
(GC)~\cite{Meade:2009iu,Bergstrom:2009fa,Meade:2009rb,Bergstrom:2008ag},
extragalactic emissions~\cite{Kamionkowski:2008gj}, and the
CMB~\cite{Slatyer:2009yq}.  There is also model building tension for
achieving such a large cross-section.  Possible mechanisms include
non-perturbative Sommerfeld
enhancements~\cite{Sommerfeld,ArkaniHamed:2008qn}, or
a resonance~\cite{Ibe:2008ye,Griest:1990kh}.  In the latter case, a very
narrow resonance and degenerate states are required, while in the
former, either large ($\gtrsim 1$) gauge or Yukawa couplings to the
light mediator or tuned parameters are
necessary~\cite{ArkaniHamed:2008qn}.  As we discuss below, the
required large couplings conflict with a need for Yukawa interactions
that generate a DM splitting, necessary in many models to avoid
constraints from direct detection~\cite{Hall:1997ah}. Indeed, the
mechanism that generates the splitting typically opens up new
annihilation channels that can parametrically dominate at freeze out.
As a consequence, in order to achieve the correct relic abundance, the couplings
responsible for the Sommerfeld enhancement are constrained and cannot
produce a large enough enhancement.

Decaying models replace the need for a large annihilation
cross-section.  Since the DM lifetime is much longer than the age of
the Universe, its decays do not affect the attractive features of the
thermal WIMP and leave no signature on the CMB radiation.  Moreover,
constraints from the GC or subhalos are easily evaded~\cite{Nardi:2008ix}, since the
emission rate depends on one power of the DM density, $\rho$, as
opposed to the $\rho^2$ dependence in the annihilating
case.  Interestingly, the correct lifetime to
explain the anomalies, $\cO(10^{26}\unit{sec})$, is obtained if the
decays are induced by dimension-6 operators suppressed by the GUT
scale~\cite{Eichler:1989br}.  Still, it is non-trivial to construct a
decaying DM model that does not over-produce antiprotons, and many
existing models are fine tuned or have small and ad hoc parameters.
 
 In this paper we study a new and natural class of models, where DM decays into a light hidden `dark sector', 
 with gauge group $\Gdark$.   Working in the supersymmetric framework
appropriate in the context of GUTs, the dark sector has a stable mass
gap at the GeV scale, and communicates with the
supersymmetric SM (SSM) through kinetic mixing~\cite{Holdom:1985ag}.  The GeV gauge
bosons decay into light SM fermions, explaining the lack of
antiproton production~\cite{Cholis:2008vb}.  The dark sector is close in spirit to the
models discussed in~\cite{ArkaniHamed:2008qn, ArkaniHamed:2008qp}.
Nonetheless, it is more general in the sense that the DM may or may
not be charged under $\Gdark$ and/or the SM\@.   This opens the door for a wider range of
models and is potentially simpler.    
Dimension-6 decay operators appear naturally, and are expected to be present at low energy unless
forbidden by global symmetries.  For related work where DM decays into light states, see~\cite{Chen:2009iua}.

 Models of the type studied here involve several scales.  Physics at the GUT
scale, $\mgut$, is responsible for producing the decay operators.  More formally, in the limit $\mgut\rightarrow \infty$, the DM is completely stable due to a preserved global symmetry.  Fields at the GUT scale then break that symmetry, inducing the required decays. 
It is important that dimension-5 operators which
would trigger a fast DM decay are not generated.   Below we show
several mechanisms that prevent such operators from showing up at low
energy.  The TeV scale generates the DM
mass which can be naturally related to the supersymmetry (SUSY) breaking scale,
thereby avoiding the usual $\mu$-problem.  The GeV scale which
controls the branching fractions of the DM decays into SM fields, is generated
either by communicating supersymmetry
breaking to the dark sector indirectly through the SM~\cite{ArkaniHamed:2008qp} or through
D-term mixing~\cite{Baumgart:2009tn,Cui:2009xq,Cheung:2009qd}.
Finally, splittings between DM states may be required to avoid direct
detection.   Such splittings are naturally of order an MeV, thereby
accommodating the inelastic DM (iDM)~\cite{TuckerSmith:2001hy} and eXciting DM (XDM)
scenarios~\cite{Finkbeiner:2007kk}.   Below we
study mechanisms that can appear at each of these scales, stressing the modular nature of such
models, which significantly simplifies the model building.   

The models studied here predict distinctive signatures in many
upcoming experiments, and unique indirect signals which  will complement the direct production experiments mentioned above.  For instance, if the dark sector is approximately
supersymmetric, or if the dark gaugino is lighter than the dark gauge
boson, $\mgaugino \lesssim \mdark$, it typically decays into a
gravitino and a SM photon.  Such primary photons will show up as sharp
features in the measured flux.  If the DM is also charged under the
SM, its decays are accompanied by primary neutrinos, again
admitting a sharp and hard spectral feature.  In the corresponding
annihilating models, these decay channels are excluded due to the
excess of primary photons or neutrinos produced, for example, at the
GC\@.  Both possibilities are studied in~\cite{Ruderman:2009ta}, where
it was shown that current and future experiments will have the ability
to measure these signatures and thereby differentiate between the
annihilating and decaying DM scenarios.  In
sections~\ref{sec:modeling-dark-sector-1} and~\ref{sec:Models} we
provide detailed examples that illustrate the presence of these
signatures.

The paper is organized as follows.  In
section~\ref{sec:modeling-dark-sector-1} we discuss the tools for
constructing decaying DM models.  We first list the dangerous pitfalls
of these models, and then discuss solutions, organized by
energy scale.  In section~\ref{sec:Models} we apply these tools to study four distinct example
models.  In~\ref{sec:U1} we show the
simplest $U(1)$ model, which is UV completed in~\ref{sec:SU2}.
In \ref{sec:SU5} we construct a model where the DM is charged under the SM and
decays into primary neutrinos, and in section~\ref{sec:u1u1} we
demonstrate how one can evade direct detection without splitting the
DM multiplets.  In section~\ref{sec:cosmology} we discuss the
cosmology of these models.  In particular, we show that a supersymmetric dark
sector can have long lived gauginos which decay into photons, without
violating constraints from big bang nucleosynthesis (BBN).  We
conclude in section~\ref{sec:conclusions}.   In
appendix~\ref{app:ModelCharges} we revisit the symmetries of the four models,
showing that these forbid the presence of any dangerous operators. 

\section{Tools for Modeling Decaying Dark Matter}
\label{sec:modeling-dark-sector-1}

In this section we describe our strategy for building models of hidden
sector decaying DM\@.  After briefly introducing our framework and
notations, we list several potential dangers for models of this type,
which arise from cosmological and experimental constraints.  We then introduce a series of model building tools, organized by energy scale, that address these dangers and can be used to build viable models.  We stress that these tools are modular, and can be used to construct a variety of models.  We demonstrate the use of these tools to build some example models in section \ref{sec:Models}.

\subsection{Framework}

\begin{figure}
\begin{center}
\includegraphics[scale=0.75]{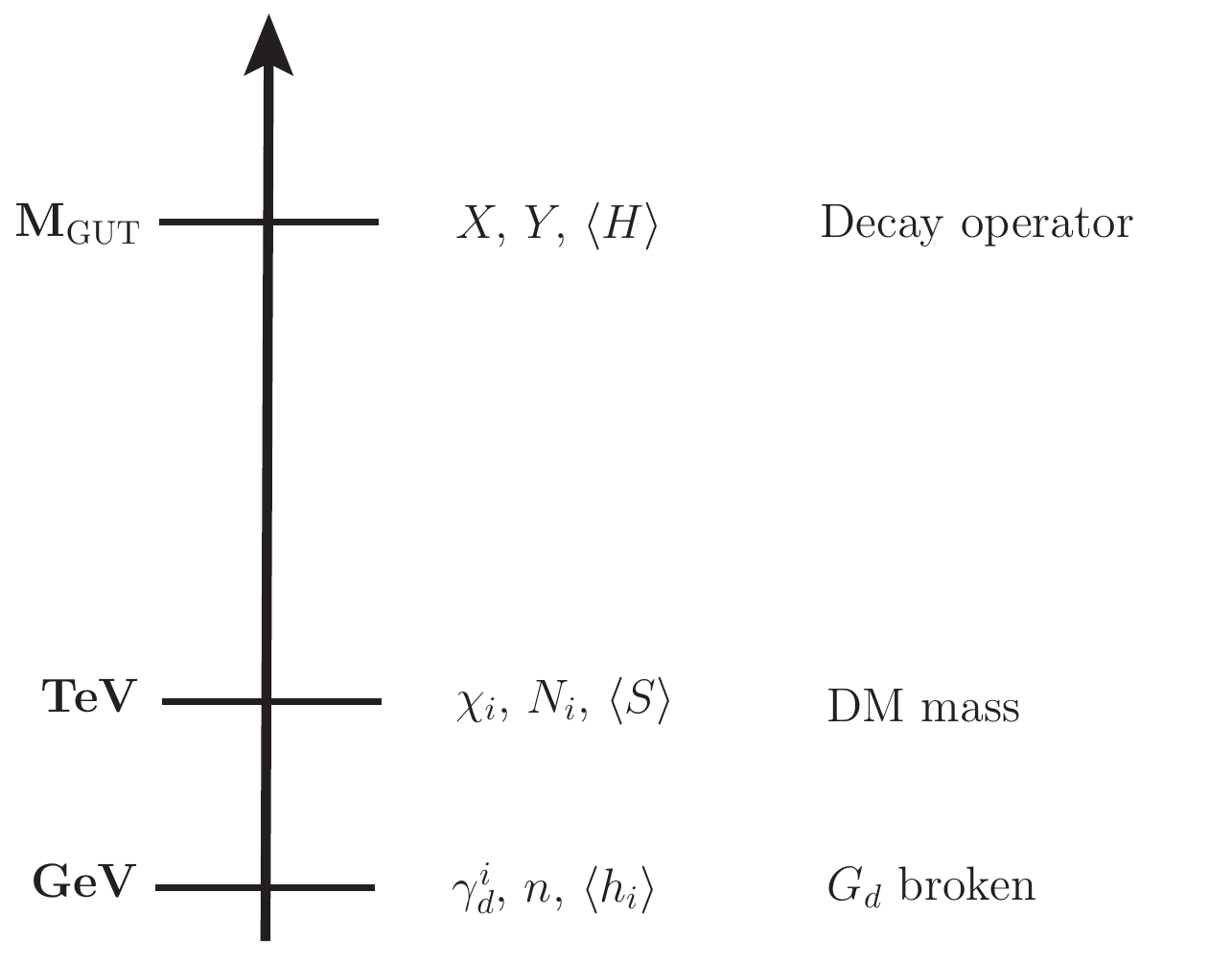}
\end{center}
\caption{ \label{fig:notation}
We summarize our notations, organized by energy scale.  $X$ and $Y$
denote GUT scale fields that are integrated out to generate
dimension-6 operators that induce DM decays.  We use $\langle H
\rangle$ to denote a GUT scale VEV, which can partially break the dark
gauge group, $\Gdark^\prime \rightarrow \Gdark$, as demonstrated in
section \ref{sec:SU2}. The DM may be composed of multiple species,
$\chi_i$, with mass at the TeV scale.  This scale is naturally
generated through the VEV of a singlet, $S$, that communicates with the SUSY breaking sector.  Here, $N_i$ denote electroweak scale fields that participate in the mechanism that generates a DM mass splitting.  Such splittings can help evade the bounds from direct detection, as we discuss in section \ref{sec:mev-scale}.  The dark gauge group, with gauge bosons $\gamma_d^i$, is entirely broken at the GeV scale by the VEVs of light Higgses.  DM decays by dimension-6 operators into these GeV scale states.  We use $h_i$ to denote light fields charged under the dark gauge group, at least some of which will receive VEVs, and we use $n$ to denote a light singlet.}
\end{figure}

We consider models where weak-scale DM, $\chi$, decays into a hidden
sector with a 
gauge group, $\Gdark$, through  a dimension-6 operator suppressed by
the GUT scale, $\mgut$.  We take this `dark sector' to be weakly
coupled with a GeV mass gap, in resemblance  to the annihilating models proposed in Ref.~\cite{ArkaniHamed:2008qn}.  Throughout this paper we work in the
supersymmetric framework which comes naturally with GUT models, and
can stabilize the GeV scale.   Furthermore, we assume the breaking of supersymmetry to be mediated through gauge interactions, allowing for a low scale of mediation.  This assumption can be somewhat relaxed, if the breaking is sequestered from the dark sector~\cite{Katz:2009qq}.  The dark sector consists of massive
gauge bosons, $\dgauge^i$, gauginos, $\dgaugino^i$, Higgses, $h_i$,
and Higgsinos, $\tilde h_i$.  We couple it to the SM
through gauge kinetic mixing, and consequently dark sector particles
decay through the mixing to SM particles. Due to the low dark gauge boson
mass, it decays predominantly into light leptons.   Within this framework, DM decays can naturally explain the PAMELA and FERMI measurements.  Our notations are summarized in Fig.~\ref{fig:notation}.

\subsection{Model Building Dangers}
\label{sec:dangers}

\begin{itemize}
\item {\bf Dimension-5 DM Decay}
\newline As we discuss in section \ref{sec:gut-scale},  dimension-6
decay operators suppressed by the GUT scale induce DM decays with a lifetime of $\tau_{6} \simeq 10^{26}$ sec, the correct timescale to account for the PAMELA and FERMI signals.  Alternatively, dimension-5 operators suppressed by the GUT scale correspond to a lifetime of $\tau_{5} \simeq 1$ sec, and must be avoided.

\item {\bf Sommerfeld Enhancement}
\newline If DM is directly charged under the light dark sector, the
annihilation cross-section is Sommerfeld-enhanced~\cite{ArkaniHamed:2008qn}.  It is important that this enhancement is
not too large, since there are various strong constraints on the
annihilation rate.  These include constraints from gamma rays and
neutrinos from the Galactic Center and Galactic Ridge
(GR)~\cite{Meade:2009rb,
  Meade:2009iu,Bergstrom:2008ag}, diffuse
gammas from extragalactic DM annihilations~\cite{Kamionkowski:2008gj},
and modified CMB radiation from DM annihilation during recombination~\cite{Slatyer:2009yq}.

\item {\bf Direct Detection}
\newline There are strong limits from direct detection on models in
which a
weak-scale DM couples elastically to a light gauge boson that
kinetically mixes with the photon.  One finds a DM-nucleon
cross-section of the order~\cite{ArkaniHamed:2008qn}:
\begin{equation}
\label{eq:directdetect}
\sigma_0 \simeq 10^{-37} \unit{cm}^2 \left( \frac{\epsilon}{10^{-3}} \right)^2 \left(
  \frac{\alpha_d}{0.01} \right) \left( \frac{\mdark}{1\unit{GeV}}
\right)^{-4}, 
\end{equation}
where $\epsilon$ parametrizes the size of the kinetic mixing.  Current
measurements rule out a cross-section of this size by 6 orders of magnitude~\cite{Ahmed:2008eu}.  There are also strong limits from direct detection on DM that couples elastically to the $Z$.  For example, models where DM is the neutral component of an $SU(2)_W$ doublet are excluded by 2-3 orders of magnitude~\cite{Cirelli:2005uq}.

\item {\bf Inelastic Capture in the Sun}
\newline As we discuss in section~\ref{sec:mev-scale}, one way to
avoid the above constraints from direct detection is to split the mass
between  the DM states,
$\delta M_{\rm DM} \gtrsim 100$ keV, and
couple inelastically to $\dgauge$ or $Z$~\cite{Hall:1997ah}.  It was recently
demonstrated that if $\delta M_{\rm DM} \simeq 100 - 500$ keV, there
are strong constraints on the inelastic capture of DM in the sun which
is followed by annihilations into $W^+ W^-$, $Z Z$, $\tau^+ \tau^-$, or $t \bar t$~\cite{Nussinov:2009ft}.  This constraint is particularly important if DM is charged under $SU(2)_W$.

\item {\bf Long-Lived GeV Scale Fields}
\newline The dark sector may contain light long-lived fields, and one
must make sure that their cosmology is safe.  On the one hand, stable
particles must not overclose the universe, $\Omega_x h^2 < 0.1$.  On
the other hand, the dark sector may contain long-lived particles that
decay electromagnetically through the kinetic mixing.  For such
decays, lifetimes of order $\tau \simeq 10^4 - 10^{12}$ sec are constrained by Big Bang Nucleosynthesis~\cite{Jedamzik:2006xz} and decays after recombination, $\tau \gtrsim 10^{13}$ sec, are constrained by diffuse gamma rays~\cite{Kribs:1996ac}. 

\item {\bf Long-Lived Colored Particles}
\newline If the DM is charged under the GUT gauge group, then there is
a colored component $\chi_{\bf 3}$.  There are strong constraints on
colored particles with lifetimes $\tau \gtrsim 10^{17}$ sec as they
form exotic atoms~\cite{Hemmick:1989ns}.  $\chi_{\bf
  3}$ must therefore have a much shorter lifetime than $\chi$.
\end{itemize}

\subsection{GUT Scale: Decay Operators}
\label{sec:gut-scale}
We consider models where weak-scale DM decays through dimension-6
operators suppressed by the GUT scale, into the dark sector.  The
GeV-scale dark fields then decay through gauge kinetic mixing to
leptons.  We focus on two possible scenarios, both of which include multiple, non-degenerate DM states: (i) One of the TeV fields receives a VEV, breaking part of $\Gdark$ at the weak scale, and (ii) none of the states obtain VEVs and $\Gdark$ is fully broken at the GeV scale.  For
scenario (ii), transitions between the TeV fields can be induced by the three body decay operators,
\begin{eqnarray}
  \label{eq:decayop3bod}
  \frac{1}{\mgut^2}\int d^4 \theta \ \chi_1^\dagger\chi_2 h_1^\dagger h_2\ , \qquad 
  \frac{1}{\mgut^2}\int d^2\theta\ \chi_1\bar \chi_2 \Wdark^2 \ , \qquad 
  \frac{1}{\mgut^2}\int d^2\theta\ \chi {\bf \bar 5}_f \Wdark^2.
\end{eqnarray}
For the first two operators, $\chi_1$ and $\chi_2$ are both weak-scale
with $m_{\chi_2} > m_{\chi_1}$.  Consequently, the DM is dominantly
composed of $\chi_2$ which generically has a larger density than
$\chi_1$.  For the third operator $\chi$ is a ${\bf 5}$ of $\SUfive$.
We will consider examples that generate each of these operators in section \ref{sec:Models}.  The decay rate of these operators is given parametrically by: 
\begin{equation} 
\label{eq:decayrate}
 \tau \simeq \left( \frac{M_{\rm
      DM}^5}{16 \pi^2 M_{\rm GUT}^4} \right)^{-1} \simeq 10^{26} \unit{sec}
\left( \frac{M_{\rm DM}}{1 \unit{TeV}} \right)^{-5} \left( \frac{M_{\rm
      GUT}}{5 \times 10^{15} \unit{GeV}}\right)^4.
\end{equation}
This is the correct timescale to account for the PAMELA and FERMI
signals, as was first noticed by Ref.~\cite{Eichler:1989br}.  For scenario (i), two body decays will typically dominate.  An example operator that we will consider in section \ref{sec:u1u1} follows from inserting a $\langle \chi_1 \rangle$ VEV into the second operator of Eq.~\eqref{eq:decayop3bod}, 
\begin{equation}
\label{eq:decayop2bod}
 \frac{1}{\mgut^2}\int d^2\theta\ \langle  \chi_1 \rangle \, \bar \chi_2 \Wdark^2 \, .
\end{equation}

As mentioned in the introduction, in the $\mgut\rightarrow \infty$ limit, the DM is completely stable.  This is typically achieved by a $\Z_2^i$ discrete symmetry under which $\chi_i$ and $\bar\chi_i$ are charged.  The superpotential at the GUT scale breaks this symmetry, destabilizing the DM\@.  We demonstrate the existence of these symmetries in the models of section~\ref{sec:Models}.  Still, such symmetries do not ensure that the DM is sufficiently long-lived.  Indeed, when integrating out GUT fields to generate the above dimension-6
decays, it is important to make sure that no dimension-5 decay
operators are generated.  This can follow from symmetries at the GUT
scale.  For each specific model of section~\ref{sec:Models} we
identify these symmetries in appendix~\ref{app:ModelCharges}.  To
demonstrate that this is possible, we now discuss two general
mechanisms for generating dimension-6 decays that do not generate
dimension-5 decays.  One simple possibility is that the hidden sector
gauge group is broken at the GUT scale, $\Gdark^\prime \rightarrow
\Gdark$, without breaking supersymmetry.  By going to Unitary gauge and integrating
out the massive $\Gdark^\prime / \Gdark$ vector superfields, it is
simple to check that dimension-6 decay operators, of the form of the
first operator in Eq.~\eqref{eq:decayop3bod}, are generated in the
\kahler potential~\cite{Seiberg:2008qj}.  Moreover, if the DM and light Higgses have
a canonical \kahler potential at the GUT scale, no dimension-5 terms are
generated.  We will discuss this in more detail for a specific example in section~\ref{sec:SU2}.

A second way to generate dimension-6 operators without generating
dimension-5 ones is by coupling canonical GUT-scale fields to the DM
in a chiral manner,
\begin{equation}
W \supset \mgut X \bar X + X \chi h\,.
\end{equation}
Integrating out $X$ and $\bar X$, and allowing for a weak-scale VEV
for the DM, results in the dimension-6 \kahler potential operator in
Eq.~\eqref{eq:decayop2bod}.  It is straightforward to see from the
equations of motion that no dimension-5 operators are generated in the
superpotential or \kahler potential.  More generally, global symmetries prevent quantum corrections from generating dimension-5 decays in the \kahler potential, as we discuss in appendix~\ref{app:ModelCharges}.

When the DM is charged under $\SUfive$, as for the third operator of
Eq.~\eqref{eq:decayop3bod}, one must ensure that its colored partner
decays on a timescale shorter than the current age of the universe.
For example, suppose that the DM is the neutral component of the
doublet of a ${\bf 5} + {\bf \bar 5}$.  The model is viable if the
triplet can decay through a dimension-5 operator that does not induce
DM decays.  This is straightforward to achieve since the triplet is
typically heavier than the doublet at the weak scale, due to the RG evolution of their masses.  Example triplet decay operators include:
\begin {equation}
\label{eq:tripletdecay}
\frac{1}{M_{\rm GUT}} \int d^2 \theta \, \chi^2 {\bf \bar 5}_f^2 \, ,\quad \quad 
\frac{1}{M_{\rm GUT}} \int d^2 \theta \, \chi {\bf 10}_f^2 s \, ,\quad \quad 
\frac{1}{M_{\rm GUT}} \int d^4 \theta\, \bar \chi {\bf \bar 5}_f^\dagger s  \, ,
\end{equation}
where in the first operator the triplet partner decays into the DM
while in the other two the triplet decays into a singlet, $s$, with
$m_{\chi_{\bf 2}} < m_s < m_{\chi_{\bf 3}}$.

\begin{figure}
\begin{center}
\includegraphics[scale=0.5]{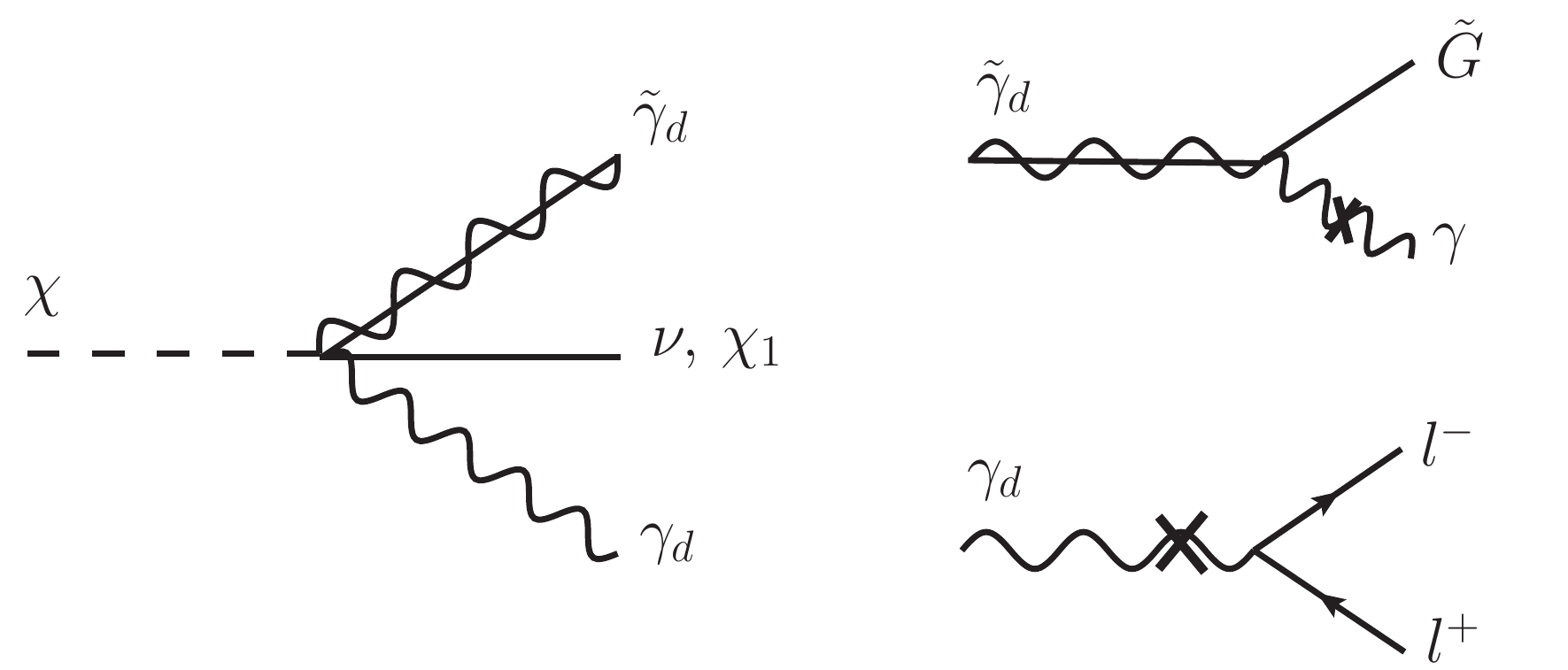}
\end{center}
\caption{A sample DM 3-body decay induced by one of the two last
  operators of Eq.~\eqref{eq:decayop3bod}.  The DM decays to a GeV-scale
  gauge boson, gaugino, and a neutrino or the lighter field, $\chi_1$.
  The gauge boson decays through the kinetic mixing to a lepton pair
  and the gaugino decays through the kinetic mixing to a photon and
  gravitino, assuming that the gaugino is lighter than, or degenerate
  with, the dark photon.  The resulting leptons can explain the PAMELA
  and FERMI excesses while the gamma rays and neutrinos lead to hard and sharp spectral features that can be probed by upcoming experiments~\cite{Ruderman:2009ta}.
\label{fig:decay}}
\end{figure}

\subsection{Weak Scale: Dark Matter Mass and Communicating SUSY Breaking}
\label{sec:weak-scale}
As we
discuss in the sections~\ref{sec:gev-scale} and~\ref{sec:conclusions}, sharp
spectral features in the photon flux may exist, depending on the light dark
spectrum.   As a consequence, the low lying excitations, and
indirectly the SUSY-breaking effects in that sector, may be probed
in the near future~\cite{Ruderman:2009ta}.   Below, we briefly discuss the possible 
effects which may influence the spectrum.

In our framework, the DM has a weak-scale mass.  A GUT-scale
one can be avoided by imposing a PQ or R symmetry that is spontaneously broken at
the weak scale by a neutral scalar, $S$.  The DM mass term then takes the form,
\begin{eqnarray}
  \label{eq:2}
   y\langle S\rangle\chi\bar\chi\,.
\end{eqnarray}
This is similar to the  well-known $\mu$-problem, and
we present no new solution.  Instead, we simply assume the coupling
above, with a VEV induced by the SUSY-breaking sector.  In principle,
$S$ may have a soft mass which arises from coupling to
the SUSY-breaking sector.  We distinguish between two cases,
\begin{itemize}
\item {\bf $\chi$ is not charged under $\Gdark$} and couples to the
  light sector only through GUT suppressed couplings.  Examples of
  such a scenario are given in sections~\ref{sec:SU5}
  and~\ref{sec:u1u1}.  In this case, SUSY-breaking effects are
  primarily communicated to the light sector through the kinetic
  mixing, as is worked out in~\cite{Morrissey:2009ur}.  The
    leading contribution to the soft mass squared of the light Higgses
    is generated as a threshold effect at the gauge messenger scale and is
    proportional to $\epsilon^2$,
 \begin{equation}
   \label{eq:msusybreak}
   \delta m_h^2 \ \simeq \  \epsilon^2\ \frac{ \gdark^2}{g_Y^2}
   M_{\tilde E}^2 \ =  \ \left(100 \unit{MeV}\right)^2 \left(
     \frac{\epsilon}{5\times10^{-4}} \right)^2 \left(
     \frac{\gdark}{g_Y} \right)^2 \left( \frac{M_{\tilde E}}{200
       \unit{GeV}} \right)^2 \,.
 \end{equation}
Here $g_Y$ is the hypercharge gauge coupling and $M_{\tilde E}$ is
 the soft mass of the right-handed selectron.    As we show in the
 next subsection, this is
 parametrically smaller by one power of $\epsilon$ compared to the
 supersymmetric mass squared of the dark vector boson.
 The corresponding contribution to the gaugino soft masses is even
 smaller~\cite{Morrissey:2009ur} and may be neglected.  The GeV scale,
 which we discuss below, is therefore approximately supersymmetric. 

\item {\bf $\chi$ is charged under $\Gdark$} and may couple directly
  to the light Higgses. Such examples are given in
  sections~\ref{sec:U1} and~\ref{sec:SU2}.  Here the SUSY-breaking
  effects can be communicated either through $S$ or through the
  kinetic mixing as discussed above.  Below, for simplicity we assume
  the latter.  We stress that $S$ can be naturally supersymmetric and
  still solve the $\mu$-problem.  This can be achieved for example
  through retrofitting~\cite{Dine:2006gm}.  We postpone the details of
  such a scenario to future work.  If, on the other hand, $S$ is
  accompanied by a soft mass, SUSY breaking effects in the light
  sector are expected to be of order GeV, and therefore dominate over
  the kinetic mixing contributions.
\end{itemize}

In the above discussion we assumed the
absence of TeV-scale messengers that couple to both the dark sector
and SM\@.  If such states exist, SUSY breaking is mediated as in gauge
mediation and the supermultiplets are split at the GeV
scale~\cite{ArkaniHamed:2008qp}.  Finally, we note that when the DM is approximately supersymmetric, both the fermion and boson components are cosmologically long-lived and constitute order one fractions of the DM relic density.  On the other hand, when there is large splitting within the DM supermultiplet, either the fermion or scalar component dominates the relic density, in a model-dependent fashion.  The analysis that follows does not depend on the spin of the DM\@.

 \subsection{GeV Scale: Breaking the Dark Sector}
\label{sec:gev-scale}

In correspondence to the discussion above, there are two ways to
naturally generate the GeV scale in the dark sector.  One is with the
use of D-term mixing which results from supersymmetric kinetic
mixing~\cite{Dienes:1996zr}.  In such a case, the light dark sector is
approximately supersymmetric, at the GeV scale.  We review this
mechanism below.  The second way to generate the GeV scale is by
communicating weak scale SUSY breaking as mentioned above.  For
simplicity, below we only consider a $\Udark$ model with the D-term mixing mechanism.
The approximate supersymmetry in the light dark sector simplifies the
analysis, since we do not need to consider GeV-scale soft terms.
Nevertheless, we stress that this is only a simplifying assumption,
which can easily be relaxed.  Indeed, introducing GeV SUSY-breaking
may change the low energy spectrum and consequently the astrophysical
signatures, but does not affect the discussions below in a significant
way.

The GeV scale of our theory resembles that
of~\cite{ArkaniHamed:2008qn,ArkaniHamed:2008qp}.  We assume that the
SM and dark sector interact with each other through gauge kinetic
mixing.  The kinetic mixing between $\Udark$ and hypercharge is given by:
\begin{eqnarray}
  \label{eq:kineticmix}
  -\frac{\epsilon}{2}\int d^2\theta\  \Wdark \W_Y.
\end{eqnarray}
$\epsilon$ is naturally of order $10^{-3} -
10^{-4}$ and arises from integrating out heavy fields charged under both sectors.
Supersymmetric kinetic mixing of this size automatically generates the
GeV scale in the dark sector~\cite{Baumgart:2009tn,Cheung:2009qd}.  To
see this, we expand Eq.~\eqref{eq:kineticmix} in components.  One
finds D-term mixing, $V \supset \epsilon D_{\rm dark} D_Y$, which upon
electroweak symmetry breaking generates a Fayet-Illiopolous (FI) term
for $\Udark$.  Such a term triggers the breaking of the dark sector at
the GeV scale:
\begin{equation}
\label{eq:mdark}
\mdark^2 =  \epsilon\ \gdark \left< D_Y \right> = \left (1 \unit{GeV} \right)^2 \left( \frac{\epsilon}{5 \times 10^{-4}} \right)  \left( \frac{\gdark}{0.35} \right) \left( \frac{\sqrt{\left< D_Y \right>}}{75 \unit{GeV}} \right)^2\,.
\end{equation}
As discussed in the previous section, the dark sector spectrum is
approximately supersymmetric when kinetic mixing is the only form of
low-energy communication between the two sectors.

When produced, the GeV scale particles can decay through the kinetic
mixing to SM particles.  The dark photon, $\dgauge$, decays directly
through the kinetic mixing to pairs of SM leptons, $l^+ l^-$.  If the
dark Higgs, $h$, is too light to decay to two dark photons, it decays
at one loop to lepton pairs.  Both of these decays are prompt on
galactic scales for typical values of the parameters:
\begin{eqnarray}
\label{eq:bosondecays}
\dgauge \rightarrow l^+ l^- &\quad& \tau \simeq (\epsilon^2 \alphaEM \mdark)^{-1} \ \simeq \ 10^{-16} \unit{sec} \,,  \nonumber \\
\vspace{0.1cm}
h \rightarrow l^+ l^- &\quad& \tau \simeq 4 \pi (\epsilon^4 \alphaEM^2 m_h)^{-1}\  \simeq \ 10^{-6} \unit{sec}\,,
\end{eqnarray}
where for the last step we have chosen the representative values
$\mdark, m_h = 1$ GeV and $\epsilon = 5\times10^{-4}$.

The decay of the lightest fermion in the dark sector has important
consequences for the astrophysical signals of our model. If the
lightest fermion mixes with the dark gaugino, it can always decay
through the kinetic mixing to the SM photon and the
gravitino\footnote{We only consider models where gravity mediation is
  not the dominant source of scale generation in the dark sector, such
  that $m_{\tilde G} \simeq F / M_p < $ GeV\@.  This is the case for the
  general framework of low-scale gauge mediation.}, $\dgaugino
\rightarrow \gamma \, \tilde G$.  The lifetime is found to be:
\begin{equation}
\label{eq:decaytophoton}
\tau_{\dgaugino \rightarrow \gamma \tilde G} \simeq  \epsilon^{-2} \left(\frac{\mgaugino^5}{16 \pi F^2} \right)^{-1} = 10^4 \unit{sec}
\left( \frac{5\times10^{-4}} {\epsilon} \right)^2 \left(\frac{1\unit{GeV}}{m_{\dgaugino}} \right)^5 \left(\frac{\sqrt{F}}{100\unit{TeV}} \right)^4\,.
\end{equation}
This decay is prompt on galactic scales for low-scale SUSY breaking,
and leads to a hard gamma ray signature.  If the lightest fermion is significantly heavier than its bosonic superpartner, it can also decay to its superpartner and the gravitino, $\dgaugino \rightarrow \dgauge \, \tilde G$, or $\tilde h \rightarrow h \, \tilde G$, with lifetime:
\begin{eqnarray}
\label{eq:decaytodarkphoton}
\tau_{\dgaugino \rightarrow \dgauge \tilde G} &\simeq& \left(\frac{m_{\dgaugino}^5}{16 \pi F^2} \right)^{-1} \left(1-\frac{\mdark^2}{\mgaugino^2} \right)^{-4} \nonumber \\
&=& 3\times10^{-3} \unit{sec}
\left(\frac{1\unit{GeV}}{m_{\dgaugino}} \right)^5 \left(\frac{\sqrt{F}}{100\unit{TeV}} \right)^4 \left(1-\frac{\mdark^2}{\mgaugino} \right)^{-4}\,.
\end{eqnarray}
 
Due to the phase space suppression above, the decay into the dark
photon is subdominant when the dark sector is approximately
supersymmetric as in our case.  Consequently DM decays into $\dgauge$ and $h$ produce hard leptons, while decays into
$\dgaugino$ produce hard gamma rays.  Since the DM decays into both bosonic and fermionic states
in the dark sector, we are led to the generic conclusion that the
hard lepton signals may be correlated with hard gamma ray signals.
These signatures are studied in detail in~\cite{Ruderman:2009ta}.

The dark spectrum and lifetimes are constrained by the requirement that the GeV scale cosmology is safe.  We discuss the dark sector cosmology and the resulting constraints in section \ref{sec:cosmology}.

\subsection{MeV Scale: Dark Matter Splitting}
\label{sec:mev-scale}
It is important for DM to evade the strong constraints on direct
detection mentioned in section~\ref{sec:dangers}.  There are three
possible solutions:
\begin{enumerate}
\item Very small kinetic mixing, $\epsilon$, between the dark sector
  and the SM\@.
\item The DM does not directly couple to $\dgauge$ or $Z$.
\item The DM multiplets are split.
\end{enumerate}
The first solution applies when DM is charged under the light dark
sector.   As we can see from equation~\eqref{eq:directdetect}, the
DM evades direct detection if the kinetic mixing is small enough,
$\epsilon \lesssim 10^{-6}$.  Interestingly, as discussed in
section~\ref{sec:cosmology}, mixing of this size may be insufficient
to keep the dark sector in thermal equilibrium thereby interfering
with the usual WIMP cosmology.

The second solution can be realized by keeping the DM neutral under
both the SM and the light gauge group.  For example, in section~\ref{sec:u1u1}, we consider a $\Utev \times \Udark$ model where DM is charged only under $\Utev$, which is broken at the weak scale, while $\Udark$ is broken at the GeV scale.  Kinetic equilibrium is maintained between DM and the SM through double kinetic mixing, as we discuss in section~\ref{sec:cosmology}.

The third possibility is to introduce a DM splitting $\delta M_{\chi}
\gtrsim 100 \unit{keV}$ .  Indeed in such a case the DM couples
inelastically together with an excited state, $\chi^\prime$, to the
dark gauge boson, $\dgauge$, or $Z$, suppressing direct
detection~\cite{Hall:1997ah}.  This bound is all that is necessary to
evade the current constraints, but there are two special values for
the splitting that are of experimental significance.  If the splitting
is of order $100$ keV, the DAMA signal~\cite{Bernabei:2008yi} can be reconciled with the
bounds from other experiments through the inelastic DM scenario
(iDM)~\cite{TuckerSmith:2001hy} (see
however~\cite{SchmidtHoberg:2009gn}).  If, on the other hand, the
splitting is of size $\delta M_{\chi} \gtrsim 1 \unit{MeV}$, it can
account for the anomalous production of positrons observed by the
INTEGRAL satellite close to the Galactic Center~\cite{Weidenspointner:2006nua}.  This is the eXciting
DM (XDM) proposal~\cite{Finkbeiner:2007kk} (see however~\cite{Lingenfelter:2009kx}).  If there are enough DM states, both
scenarios can be realized.

Suppose first, that DM is charged under $\Gdark$.  Splittings with the right parametric size for iDM or XDM are generated by direct couplings between DM and the light Higgses~\cite{Cheung:2009qd}:
\begin{equation}
\label{eq:split}
W \supset S \left(y_N N^2 + y_\chi \chi \bar \chi \right) + y_{\rm
  split} N \chi h \,.
\end{equation}
As discussed above, we assume that $S$ interacts with the SUSY breaking sector and gets a weak scale VEV\@.  $N$ is a singlet and stability of DM requires $N$ to be heavier than $\chi$, $\left| y_N \right| > \left| y_\chi \right|$.  In this limit, we integrate out $N$ and find a DM splitting of size: 
\begin{equation}
\label{eq:splitsize}
\delta \mdm = \frac{y_{\rm split}^2 \left< \bar h \right >^2}{4 m_N} =
100 \unit {keV} \left( \frac{y_{\rm split}}{1} \right)^2
\left(\frac{\left< \bar h \right>}{1 \unit{GeV}} \right)^2 \left(
  \frac{m_N}{2.5 \unit {TeV}} \right)^{-1} \,.
\end{equation}
 If $\chi$ is charged under the SM, the last term in
 Eq.~(\ref{eq:split}) can be replaced with a coupling to the SM Higgs.
 In that case the splitting is expected to be larger.
 
There is an important caveat to the above mechanism.  If $S$ gets a
weak-scale $F$-term, the $\chi$ scalars receive a weak-scale splitting
and the dark gauge boson couples across the splitting.  This
SUSY-breaking splitting provides another mechanism for evading the
constraint from direct detection, but the splitting is generically too
large to account for iDM or XDM\@.  If we wish to include these
proposals, $S$ must receive a weak-scale VEV but should have no F-term
to leading order\footnote{An alternative possibility is to introduce
  another source of SUSY breaking that lifts both $\chi$ scalars above
  the fermions so that the fermions constitute DM\@.}.  Consequently,
$S$ cannot be the NMSSM singlet.  We note that there are more
options other than Eq.~\eqref{eq:split} for generating an MeV size DM splitting, and we will employ a slightly different mechanism in appendix~\ref{app:SU2}.

Another possibility is that the DM is charged under a
non-Abelian hidden sector.  In this case, the splittings among the DM multiplet are generated radiatively after dark sector symmetry breaking~\cite{ArkaniHamed:2008qn}.  In practice, non-Abelian dark sectors are more difficult to construct and often require elaborate Higgs sectors~\cite{Baumgart:2009tn}.  In the explicit models that we study below, we will instead focus on the simplest possibility of a $\Udark$ hidden sector at low energies.  This is for illustrative purposes only, and more complicated dark sectors remain a valid possibility.

\section{Models}
\label{sec:Models}

In this section we use the tools described above to construct four
explicit models of hidden sector decaying DM\@.  There are many possible
models within this framework, and these should be viewed only as
illustrative examples.  The models are roughly ordered by increasing
complexity.  We begin with a minimal $\Udark$ dark sector which
includes all of the main ideas.  The second model embeds $\Udark$ into
$\SUtwodark$ at the GUT scale.  The $\SUtwodark$ breaking generates
dimension 6 DM decay.  For the third model, we consider DM charged
under the SM, and find that there is always an associated hard
neutrino signal.  The reader who is primarily interested in the new
correlated signals that we propose may want to skip directly to this
model.  All four models can produce hard gammas that are correlated
with the astrophysical leptons, but only the third model also produces
hard neutrinos.  Finally, we consider a $\Utev \times \Udark$ model,
where no splitting is required to evade direct detection.  In section
\ref{sec:cosmology}, we discuss the constraints that cosmology places
on these models.  In appendix \ref{app:ModelCharges}, we discuss some
further technical details for each model.

In the last two models the DM is not charged under the GeV-scale dark
sector.  Consequently no  Sommerfeld enhancement is present at all, so
the astrophysical constraints are automatically avoided.  Such models
are in sharp contrast to the annihilating models of~\cite{ArkaniHamed:2008qn}.

\subsection{$\Udark$: The Minimal Model}
\label{sec:U1}

\begin{figure}
\begin{center}
\includegraphics[scale=0.5]{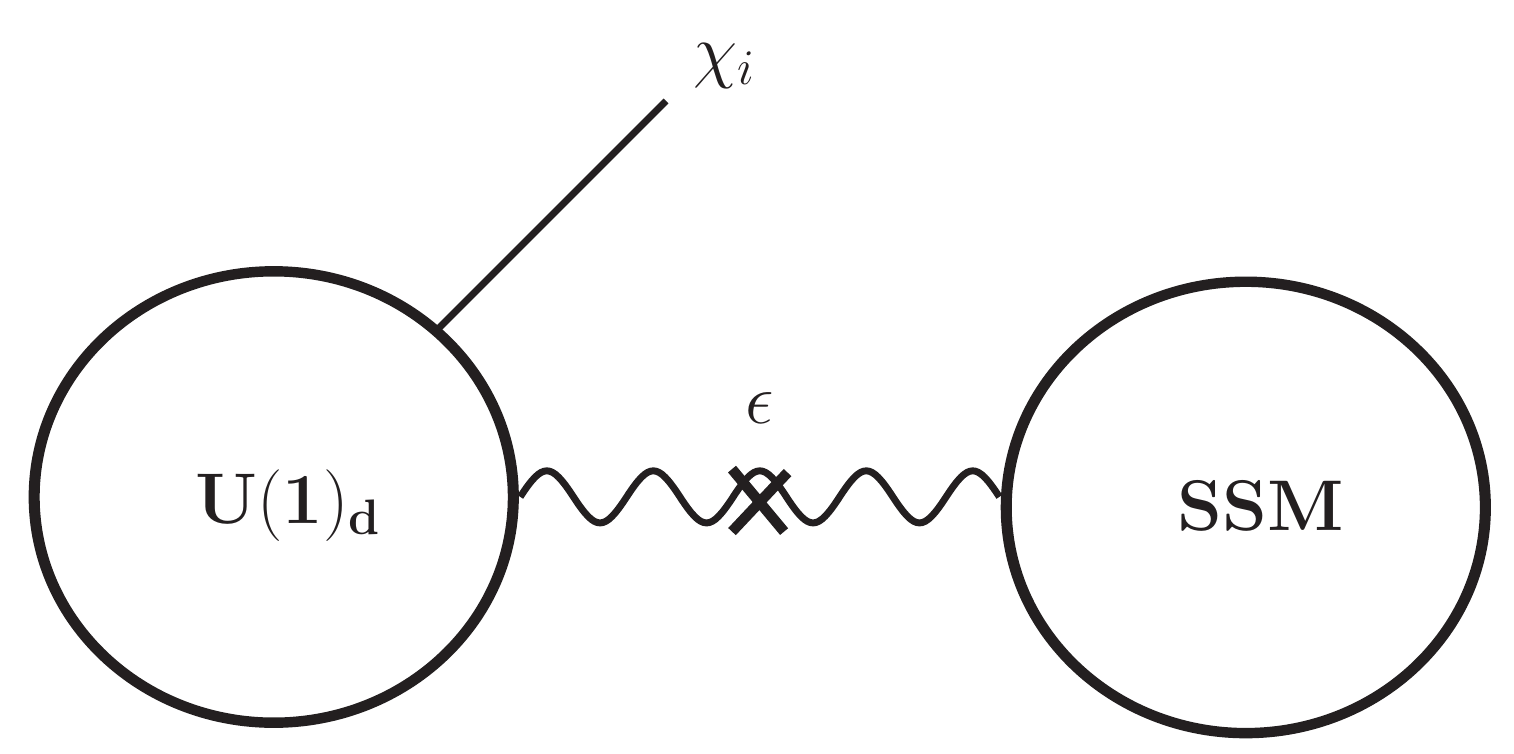}
\end{center}
\caption{The setup of our minimal model.  DM is charged under the
  hidden sector, $\Udark$, and decays through dimension-6 GUT scale
  suppressed operators into the dark sector gauge multiplet.  $\Udark$
  kinetically mixes with hypercharge, and this kinetic mixing has
  three important effects: (i) D-term mixing causes $\Udark$ to break
  at the GeV scale, (ii) dark gauge bosons decay through the kinetic
  mixing to leptons while decays into antiprotons are kinematically
  forbidden, and (iii) DM stays in kinetic equilibrium with the SM
  through the kinetic mixing, allowing for the usual `WIMP Miracle'
  cosmology (see section~\ref{sec:cosmology}).
  \label{fig:u1}}
\end{figure}

\begin{table}[t]
\vskip 0.5cm
\begin{center}
\begin{tabular}{|c|| c c c c | c c c c  | c c c |}
\hline
& \multicolumn{4}{c|}{GUT}& \multicolumn{4}{c|}{TeV}
&\multicolumn{3}{c|}{GeV} 
\\
&                                  $X$ & $\bar X$& $Y$ & $\bar Y$ &    $\chi_i$ & $\bar \chi_i$ & $S$ & $N_i$ & $h$ & $\bar h$ & $n$
\\  \hline \hline
$\Udark$ &  0  & 0  &       1 &  -1 &           1 &          -1  &                    0 &     0 &        1 & -1 &       0
\\ \hline
\end{tabular}
\end{center}
\caption{The matter content of the $\Udark$ model, where $i = 1,2$.  We stress the modularity of the model by grouping the fields according to their scales.}
\label{tab:u1}
\end{table}

We begin by considering the simplest possibility, $\Gdark = \Udark$.
This model captures the main ideas of our framework and serves as an
example for the models that follow.  We assume a kinetic mixing
between $\Udark$ and hypercharge, as in equation
\eqref{eq:kineticmix}.  The field content is listed in table
\ref{tab:u1} and the setup is illustrated in Fig.~\ref{fig:u1}.  All
fields are assumed to have  canonical \kahler potential at the GUT
scale.  In order to stress the modularity of the model, we split up
the superpotential into three pieces,
\begin{equation}
\label{eq:U1W}
W = W_{\rm decay} + W_{\rm DM} + W_{\rm split} \,.
\end{equation}
The first term leads to DM decay:
\begin{equation}
\label{eq:U1Wdecay}
W_{\rm decay} = \left(\mgut + X\right) Y \bar Y +  \mgut X\bar X + \bar X \chi_1 \bar \chi_2 \,.
\end{equation}
Integrating out the GUT scale fields generates the second operator of
equation \eqref{eq:decayop3bod}, at one loop~\cite{Arvanitaki:2008hq}.
Meanwhile, it is straightforward to see from the equations of motion
that no dimension-5 decays are generated in the superpotential.

The second term of equation \eqref{eq:U1W} determines the DM and dark sector spectrum:
\begin{equation}
\label{eq:U1Wdm}
W_{\rm DM}  = S\left( y_1 \chi_1 \bar \chi_1+ y_2 \chi_2 \bar \chi_2 \right) +
n h \bar h \,.
\end{equation}
We assume that $S$ obtains a weak-scale VEV, possibly through
interactions with the SUSY breaking sector.  The different Yukawa couplings $y_{1,2}$ generate masses for
$\chi_1$ and $\chi_2$ with $m_{\chi_2} > m_{\chi_1}$.  Both $\chi_i$
are stable on cosmological timescales and contribute to the relic
density, however, DM is mostly composed of $\chi_2$, whose larger mass
leads to a smaller annihilation cross-section and therefore to a
larger abundance.  The dimension-6 decay operator in
Eq.~(\ref{eq:decayop3bod}), leads to three body decays of $\chi_2$ into
$\chi_1$ and dark sector gauge bosons, $\dgauge$, and/or gauginos,
$\dgaugino$.  $\dgauge$ and $\dgaugino$ then decay to SM leptons and
photons through the channels described in section~\ref{sec:gev-scale}.

At the GeV scale, this model resembles the low-energy $U(1)$
construction of Ref.~\cite{Cheung:2009qd}.  The D-term mixing,
described in section~\ref{sec:gev-scale}, generates an effective FI
term for the $\Udark$, which triggers one of the light Higgses to get a VEV
at the GeV scale.  Without loss of generality, we take $\bar h$ to be
the one with a non-vanishing VEV\@.  Expanding around $\langle \bar
h\rangle$, $h$ and $n$ obtain a GeV mass
through the last term of 
Eq.~\eqref{eq:U1Wdm}.  Consequently, all fields are lifted, forming a GeV scale mass gap.

The last term of Eq.~\eqref{eq:U1W} corresponds to two copies of
the splitting mechanism described by Eqs.~\eqref{eq:split},\eqref{eq:splitsize},
\begin{equation}
\label{eq:U1Wsplit}
W_{\rm split} = \sum_{i=1}^2 \left(S N_i^2+N_i \chi_i \bar h \right)\,.
\end{equation}
Splittings are generated for both $\chi_i$, evading the constraints
from direct detection.  The two splittings are of different sizes, and
we note that both iDM and XDM can be incorporated in this model if $\delta
M_{\chi_1} \sim 100$ keV and $\delta M_{\chi_2} \sim 1$ MeV, or vica
versa.  It would be interesting to conduct a more detailed study of
this multi-species DM model to see if indeed the two scenarios can be accommodated.

That $\chi_1$ and $\chi_2$ are long-lived follows from an unbroken $\Z_2^1 \times \Z_2^2$, as $\mgut \rightarrow \infty$.  $\chi_i, \bar \chi_i$, and $N_i$ are charged under $\Z_2^i$, respectively.  This symmetry is broken by Eq.~\eqref{eq:U1Wdecay}, resulting in DM decays.  In appendix~\ref{app:ModelCharges}, we verify that dimension-5 decays are forbidden by a GUT scale symmetry.

We conclude by remarking that with this field content, the
absence of Landau poles below the GUT scale places a bound on the dark
gauge coupling at the GeV scale, $\alphadark \lesssim 1/30$.  

\subsection{$\SUtwodark \rightarrow \Udark$: GUT Scale Symmetry Breaking}
\label{sec:SU2}

\begin{table}[t]
\vskip 0.5cm
\begin{center}
\begin{tabular}{| c ||c c  c c | c c c c  | c c c |}
\hline
& \multicolumn{4}{|c|}{GUT}& \multicolumn{4}{c|}{TeV}
&\multicolumn{3}{c|}{GeV}
\\ 
 & $H$ & $X$ & $\bar \Phi$& $\bar n$&$\chi$ & $\bar \chi$ & $\Phi$ & $S_\Phi$ & $h$
 & $n^\prime$ & $s_n$
\\  \hline \hline
$\SUtwodark$ & {\bf Adj} & $\fund$&{\bf Adj}&{\bf Adj}&$\fund$&$\fund$&{\bf
  Adj}&{\bf 1}&$\fund$&{\bf Adj} & {\bf 1}
 \\
\hline 
\end{tabular}
\end{center}
\caption{The matter content for the $\SUtwodark \rightarrow \Udark$ model.}
\label{tab:su2}
\end{table}

We now consider a UV completion of the previous model, by embedding
$\Udark$ into $\SUtwodark$ which is broken at the GUT scale.  In the
following discussion, we focus on the two new features of this model:
(i) heavy gauge bosons generate the dimension-6 DM decay, and (ii) the low-energy theory contains split $\SUtwodark$ multiplets.  The field content is summarized in table \ref{tab:su2}.  Again, we assume a canonical \kahler potential and group the terms in the superpotential according to their role,
\begin{equation}
\label{eq:SU2W}
W = W_{\rm decay}+W_{\rm GUT}+W_{\rm DM} \,.
\end{equation}

The first term above, triggers the GUT-scale breaking $\SUtwodark
\rightarrow U(1)_d$,
\begin{equation}
  \label{eq:SU2Wdecay}
  W_{\rm decay} = f(H) + H X^2\,.
\end{equation}
Here $H = H^a T^a$ is a triplet and $T^a = \sigma^a/2$ are the
generators of $\SUtwodark$.  In most of what follows we suppress color
indices.  We take $f(H)$ to be a potential for $H$ with a minimum at
$\left< H \right> = \mgut T^3$.  Consequently, $X$, which is
introduced to cancel $SU(2)$ anomalies~\cite{Witten:1982fp}, obtains a
GUT-scale mass and is integrated out.

To see the effect of the breaking, we integrate out the broken
$\SUtwodark/\Udark$ generators.  Going to the Unitary gauge and
solving for the massive vector superfields, $V_{\pm} = V_1 \mp i V_2$,
one finds an additional contribution to the \kahler
potential~\cite{Seiberg:2008qj},
\begin{eqnarray}
  \label{eq:SU2K}
  \delta K_{\rm eff} = - (\varphi_i^\dagger
  T^+\varphi_i)\lambda_{\pm}^{-1}(\varphi_j^\dagger T^-\varphi_j)\,,
\end{eqnarray}
where $\varphi_i$ collectively denote all fields (subject to the
Unitary gauge constraint), $T^\pm=T^1 \pm iT^2$ are the broken
generators in the corresponding representation and
\begin{eqnarray}
  \label{eq:SU2lambda}
  \lambda^{\pm} = \frac{1}{2} H^\dagger \{ T^+, T^-\} H = \mgut^2\,.
\end{eqnarray}
Substituting the DM states, $\chi_\alpha = (\chi_1,\chi_2)$, $\bar
\chi^\alpha = (\bar\chi_1, \bar\chi_2)$ and light Higgs, $h_\alpha =
(h_1,h_2)$ into Eq.~\eqref{eq:SU2K} one finds the contributions,
\begin{eqnarray}
  \label{eq:SU2decay}
  - \frac{1}{\mgut^2} \int d^4 \theta \ \left( \chi_1^{\dagger} \chi_2
  h_2^{\dagger} h_1 + \bar\chi_1^{\dagger} \bar \chi_2 h_1^{\dagger} h_2
\right) + (1\rightarrow2)  \,.
\end{eqnarray}
These operators are precisely of the form of the first operator in
Eq.~(\ref{eq:decayop3bod}).  As in the $\Udark$ model, we will require
$m_{\chi_2} > m_{\chi_1}$.  In this case, DM is mostly composed of
$\chi_2$ and Eq.~\eqref{eq:SU2decay} generates 3-body decay of
$\chi_2$ into $\chi_1$ and the lights Higgses $h_1$ and $h_2$.  We again
assume that there is kinetic mixing between the low-energy $U(1)_d$
and hypercharge, generated by integrating out fields charged under
both the dark sector and SM, so that D-term mixing generates a
GeV-scale VEV for $h_2$, which is eaten by the gauge multiplet and
decays to SM leptons and photons as we describe in section
\ref{sec:gev-scale}.

In order to minimize the low-energy field content so that it matches
the $\Udark$ model of the previous section, and in order to give
different masses to $\chi_1$ and $\chi_2$, we work with split
$\SUtwodark$ multiplets.  We can split the triplets $\Phi$ and $n$ by
coupling them to $H$ and GUT scale singlets, denoted by $S_\Phi$ and
$s_n$:
\begin{equation}
  \label{eq:SU2Wgut}
  W_{\rm GUT} = \tr \left[g(H) \left( \Phi\bar\Phi+ S_\phi\bar\Phi +
      n^\prime \bar n +  s_n\bar n \right) \right] \,. 
\end{equation}
For generic $g(H)$\footnote{There is in general a different polynomial
  of $H$ in front of each term of Eq.~\eqref{eq:SU2Wgut}, which we
  have suppressed to keep our notation compact.}, the VEV of $H$
generates GUT scale masses for all component fields except for one
linear combination of $\Phi_3$ and $S_\Phi$, which we denote by $S$, and
one linear combination of $n^\prime_3$ and $s_n$, which we denote $n$.
The specific linear combinations that remain light depend on $g(H)$.
In appendix \ref{app:ModelCharges}, we use a discrete symmetry to
prove that $S$ and $n$ remain light and to show that dimension-5 DM
decays can be forbidden for generic superpotentials.

The low-energy theory is dictated by the superpotential terms:
\begin{equation}
\label{eq:SU2Wdm}
W_{\rm DM} = (\Phi + S_\Phi) \chi \bar \chi  + (n^\prime+s_n) h^2\,.
\end{equation}
After the $\SUtwodark$ breaking splits the multiplets, the low-energy
effective superpotential is of the form:
\begin{equation}
\label{eq:SU2Wdmeff}
W_{\rm eff} = S  (y_1\, \chi_1 \bar \chi_1 + y_2\,\chi_2 \bar \chi_2) + n h_1 h_2\,,
\end{equation}
with $y_i$ couplings of order one that depend on $g(H)$.  The
projection onto the light state $S$ results in couplings that are not
$SU(2)$ invariant, and the TeV scale VEV of $S$ therefore generates
different masses for $\chi_1$ and $\chi_2$.  As before, the DM is long-lived because as $\mgut \rightarrow \infty$, there is an unbroken $\Z_2^1 \times \Z_2^2$ symmetry, under which $\chi_i$ and $\bar \chi_i$ are separately charged for $i=1,2$.   The third term is the
same as the last term of Eq.~\eqref{eq:U1Wdm}, and the low-energy dark
sector is thus the same as the $\Udark$ model.  It is also
straightforward to induce small DM splittings, in order to evade the
constraints from direct detection and possibly incorporate iDM and
XDM\@.  This is shown in appendix \ref{app:ModelCharges}.

\subsection{$\SUfive \times \Udark$: SM Charged-DM and Correlated Neutrinos}
\label{sec:SU5}

\begin{figure}
\begin{center}
\includegraphics[scale=0.5]{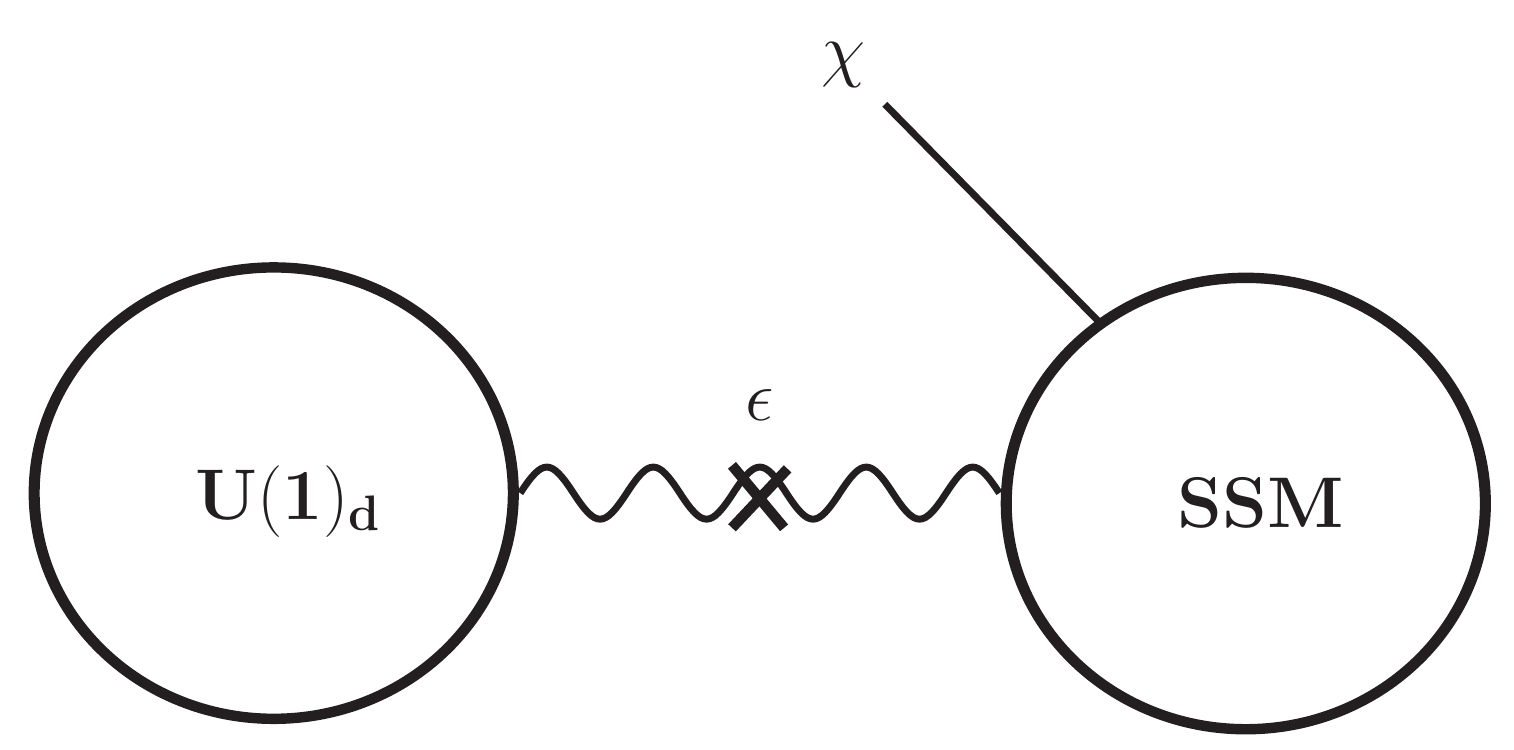}
\end{center}
\caption{A model with DM charged under the SM\@.   DM is the neutral
  component of a $\bf 5 + \bar 5$ representation of $SU(5)_{\rm SM}
  \supset SU(3)_C\times SU(2)_W \times U(1)_Y$.  DM decays through a
  dimension-6 operator into the gauge multiplet of $\Udark$, which
  kinetically mixes with hypercharge.  The conservation of hypercharge
  (at the GUT scale) implies that this decay must be accompanied by
  the associated production of a neutrino. This results in a primary neutrino spectrum that is correlated with the leptonic cosmic rays, and will be tested by upcoming experiments such as IceCube/DeepCore~\cite{Ruderman:2009ta}.
  \label{fig:su5}}
\end{figure}

\begin{table}[t]
\vskip 0.5cm
\begin{center}
\begin{tabular}{|c||c c c c | c c c c c | c c c |}
\hline
& \multicolumn{4}{|c|}{GUT}& \multicolumn{5}{c|}{TeV}
&\multicolumn{3}{c|}{GeV} 
\\
 & $X$ & $\bar X$ &$Y$&$\bar Y$& $\chi$ & $\bar\chi$ & $S$ & $s_1$ & $N$ & $h$ & $\bar h$ & $n^\prime$  
\\  \hline \hline
$\SUfive$ & {\bf 1}  & {\bf
  1} &  {\bf 1} & {\bf
  1}  &$\fund$& $\overline\fund$ & {\bf 1} & {\bf 1} & {\bf 1} &
{\bf 1} & {\bf 1} & {\bf 1}  
\\ 
$\Udark$ & 0   & 0  &  1&  -1  &  0& 0 & 0 & 0 &0& 1 & -1& 0  
\\ \hline
\end{tabular}
\end{center}
\caption{The matter content for our model with DM charged under $\SUfive \supset SU(3)_C \times SU(2)_W \times U(1)_Y$.}
\label{tab:su5}
\end{table}

We now consider a model where DM is charged under the SSM and decays
through a dimension-6 operator into the dark sector.  In this model,
DM itself is not charged under the GeV sector, avoiding the
constraints due to Sommerfeld enhancement discussed in section
\ref{sec:dangers}.  We take $\chi+\bar\chi$ to be charged under an
$SU(5)$ GUT gauge group, residing in a ${\bf 5} + {\bf \bar 5}$. A
schematic description of the model is shown in Fig.~\ref{fig:su5}.  By
gauge invariance, decay into the dark sector must be accompanied by
associated SM particle production\footnote{We thank N.~Arkani-Hamed
  for drawing our attention to this point.}.  If one SM particle is
produced, it must be a neutrino or Higgs.  The latter produces
antiprotons, which are constrained by PAMELA, and thus we focus on the
possibility that DM decays produce hard neutrinos that accompany dark
sector production.  The discovery of such neutrinos is discussed
in~\cite{Ruderman:2009ta}. 

An important requirement for this model is that the colored partner of
DM decays faster than the current age of the universe.  This is because
there are strong constraints on stable colored particles, as discussed
in section \ref{sec:dangers}.  These constraints are evaded if the
triplet DM decays through a dimension-5 operator.  For this model, we
assume that the canonical \kahler potential is supplemented by one
irrelevant operator, generated at the GUT scale,
\begin{equation}
\label{eq:SU5K}
K_{\rm DM} =  \frac{1}{\mgut} \int d^4 \theta \  \chi \, {\bf \bar 5}_f^\dagger s_1\,,
\end{equation}
where $s_1$ is a singlet with mass: $m_{\chi_{\bf 2}} < m_{s_1} <
m_{\chi_{\bf 3}}$.  This mechanism can be easily arranged since the
triplet partner is expected to be 
heavier than the DM, due to the RG evolution of their
masses below the GUT scale.   

We list the field content in table \ref{tab:su5}, and again we group
the superpotential terms according to their roles:
\begin{equation}
\label{eq:SU5W}
W = W_{\rm decay} + W_{\rm DM} + W_{\rm split}\,.
\end{equation}
The first term generates dimension-6 DM decay using the same mechanism
as our $\Udark$ model of section \ref{sec:U1}:
\begin{equation}
\label{eq:SU5Wdecay}
W_{\rm decay} =  (\mgut + X) Y \bar Y + \mgut X \bar X + \bar X \chi{\bf \bar 5}_f \,.
\end{equation}
Integrating out $X$ and $Y$ generates the third dimension-6 decay
operator of Eq.~\eqref{eq:decayop3bod} at one-loop:
\begin{equation}
\label{eq:SU5decay}
  \frac{1}{\mgut^2}\int d^2\theta\ \frac{\alphadark}{4 \pi} \chi {\bf
    \bar 5}_f \Wdark^2 \,.
\end{equation}
This operator results in three-body decay, with DM decaying into one
neutrino or sneutrino and two dark gauge bosons or gauginos, which
subsequently 
decay to SM leptons and photons through the operators discussed in
section \ref{sec:gev-scale}.

At low energies, this model resembles the constructions above:
\begin{equation}
\label{eq:SU5Wdm}
W_{\rm DM} = S \left( \chi \bar \chi + s_1^2 \right) + n h \bar h
\end{equation}
We assume that $S$, which may be the NMSSM singlet, gets a weak-scale
VEV\@.  This generates a mass for the DM and the singlet $s_1$, which plays
a role in the triplet decay of Eq.~\eqref{eq:SU5K}.  As in
the models above, we take $\Udark$ to kinetically mix with
hypercharge, and the D-term mixing generates a GeV-scale VEV for $\bar
h$.  With no DM splitting, this model would be ruled out because the DM
couples strongly to the SM $Z$ boson.  This constraint is evaded by
coupling the DM
to the Higgs, which generates a small splitting:
\begin{equation}
\label{eq:SU5Wsplit}
W_{\rm split} = S N^2 + \chi H_d N \,.
\end{equation}
Here $N$ is a singlet that must be heavier than $\chi$, to ensure its stability.
The resulting splitting is too large to account for iDM or XDM\@.  In
fact, iDM is already ruled out for this model by the constraints from
inelastic capture in the sun, as discussed in section
\ref{sec:dangers}.   Finally, the DM is long-lived due to an unbroken $\Z_2$ at the renormalizable level, under which $\chi, \bar \chi$, and $N$ are charged. 

If the DM relic density is only determined by its $SU(2)_W$ gauge
interaction, its mass is fixed to be: $m_\chi \simeq 1.1$ TeV~\cite{Cirelli:2005uq}.  This mass is too small to fit the FERMI excess
with DM decays~\cite{Meade:2009iu}.  Fortunately, the second operator
of the above splitting mechanism, Eq.~\eqref{eq:SU5Wsplit}, opens up a
new annihilation channel into SM Higgses.  This raises the DM
annihilation cross-section, allowing for heavier masses which can fit
FERMI\@.  We discuss the DM relic density further in section
\ref{sec:cosmology}.

\subsection{$\Utev \times \Udark$: No Mass Splitting}
\label{sec:u1u1}

\begin{figure}
\begin{center}
\includegraphics[scale=0.5]{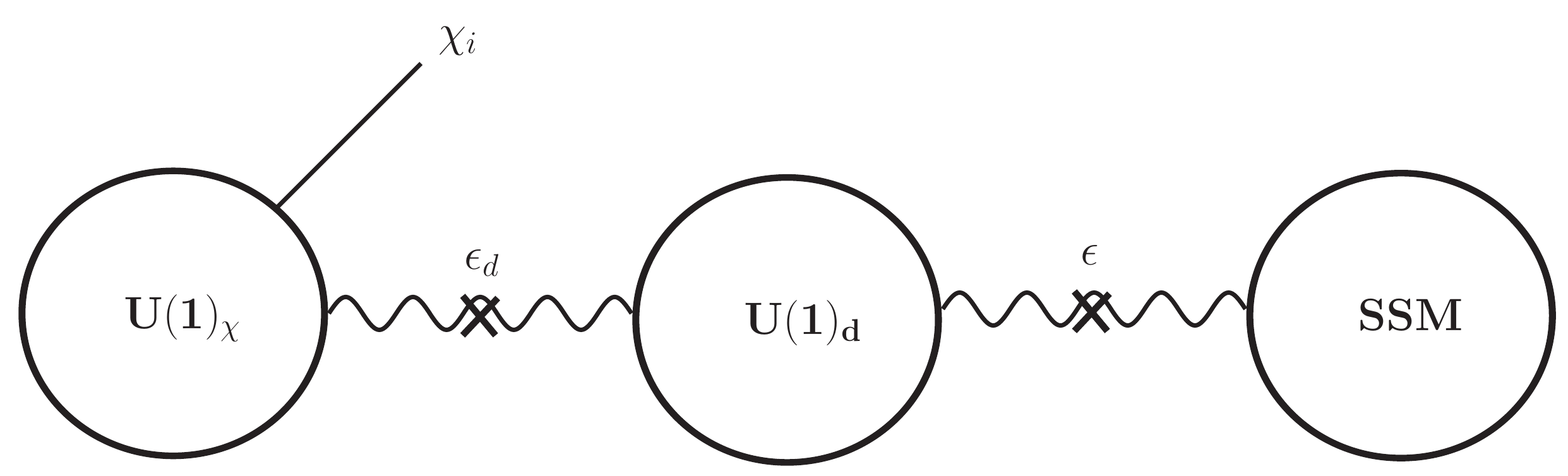}
\end{center}
\caption{A model with double kinetic mixing.  DM, $\chi_2$, is charged under
  $\Utev$, which is broken by a different species, $\chi_1$, at the TeV scale.  Decays are induced
  by a dimension-6 GUT suppressed operator into the $\Udark$ gauge multiplet, which kinetically mixes with both hypercharge and
  $\Utev$.  This double kinetic mixing is sufficient to keep DM in
  kinetic equilibrium with the SM, preserving the usual WIMP cosmology
  (see section~\ref{sec:cosmology}).  There is no strong constraint
  from direct detection because the DM does not couple directly to the $Z$ or $\Udark$ gauge boson, and therefore no DM splitting is required.
\label{fig:u1u1}}
\end{figure}

\begin{table}[t]
\vskip 0.5cm
\begin{center}
\begin{tabular}{|c||c ccc | c c c | c c c | }
\hline
& \multicolumn{4}{c|}{GUT}& \multicolumn{3}{c|}{TeV}
&\multicolumn{3}{c|}{GeV} 
\\ 
& $X$ & $\bar X$ & $Y$ & $\bar Y$     &$\chi_i$ &  $\bar\chi_i$& $S_i$ &  $h$ &        $\bar h$ & $n$  
\\ \hline \hline
$\Utev$ &  0 &  0 &  0 & 0& 1 & -1 & 0 & 0 & 0 & 0  
\\ 
$\Udark$ & 0 & 0 & 1 & -1 &0& 0  & 0 & 1 & -1 & 0 
\\ \hline 
\end{tabular}
\end{center}
\caption{The matter content for the $\Utev \times \Udark$ model, where $i = 1,2$.}
\label{tab:u1u1}
\end{table}

We now consider a model with a $\Utev \times \Udark$ hidden sector.
We illustrate the basic idea in Fig. \ref{fig:u1u1}.  The DM, $\chi_2$, is charged
under $\Utev$, which is broken at the weak scale by the VEV of a different species $\chi_1$.  It decays through a
dimension-6 operator into the $\Udark$ gauge multiplet.  There are
two advantages to this setup.  First, this model does not have a
strong constraint from direct detection, because DM does not couple
directly to the $Z$ boson or $\dgauge$.  Therefore, unlike the previous
models, no DM splitting is required.  Second, there is no constraint
from photon or neutrino measurements, as in the model of section \ref{sec:SU5},
since DM is not charged under $\Udark$.   Another unique feature of
this model is that 2-body decays dominate over 3-body decays because the decay operator contains a field, $\chi_1$, which obtains a weak scale VEV\@.

The field content of this model is listed in table \ref{tab:u1u1}.  We assume a canonical \kahler potential, and we group the superpotential terms according to their role,
\begin{equation}
\label{eq:U1U1W}
W = W_{\rm decay} + W_{\rm DM}\,.
\end{equation}
The first term is identical to the GUT scale interactions of the $\Udark$ model, Eq.~\eqref{eq:U1Wdecay},
\begin{equation}
\label{eq:U1U1Wdecay}
W_{\rm decay} = \left(M_{\rm GUT} + X\right) Y \bar Y\nonumber +  M_{\rm GUT} X \bar X + \bar X \chi_1 \bar \chi_2 \,,
\end{equation}
generating the second decay operator of Eq.~\eqref{eq:decayop3bod}.
Integrating out bifundamentals generates kinetic mixing between $\Udark$ and $\Utev$,
\begin{equation}
\label{eq:U1U1KM}
-\frac{\epsilon_d}{2} \int d^2 \theta \ \W_\chi \W_d\,,
\end{equation}
of the same size as the kinetic mixing between $\Udark$ and hypercharge, $\epsilon_d \sim \epsilon \sim 10^{-3} - 10^{-4}$.
A mixing of this size is small enough to keep
the $\Udark$ mass gap at a GeV, but large enough to keep the $\Utev$
sector in thermal equilibrium with the $\Udark$ sector.  The latter
guarantees, through the double kinetic mixing, that $\Utev$ is in
thermal equilibrium with the SM\@.  We discuss the cosmology of this
model in more detail in section \ref{sec:cosmology}.

The terms in the superpotential relevant at low energies are:
\begin{equation}
\label{eq:U1U1Wdm}
W_{\rm DM} =  S_2 \, \chi_2 \bar \chi_2  +S_1(\chi_1 \bar \chi_1 + S_2^2) + n h \bar h\,,
\end{equation}
where $S_2$ receives a weak scale VEV from communicating with the SUSY breaking sector, giving the DM a mass.  Solving for the $F$-term of $S_1$, one finds VEVs for $\chi_1$
and $\bar \chi_1$ of order $\langle S_2 \rangle$.  This breaks $\Utev$ at the weak
scale, and the dominant DM decay is 2-body, with a $\chi_1$ VEV insertion resulting in the operator of Eq.~\eqref{eq:decayop2bod}.  DM is long-lived because as $\mgut \rightarrow \infty$ there is an unbroken $\Z_2$ symmetry, $(\chi_2, \bar \chi_2) \rightarrow - (\chi_2, \bar \chi_2)$.
As in the previous models,
$\bar h$ gets a VEV at the GeV scale due to the D-term mixing between hypercharge and $\Utev$.  The last term of Eq.~\eqref{eq:U1U1Wdm} generates a GeV scale mass gap.

\section{Cosmology of the Dark Sector}
\label{sec:cosmology}

In this section we discuss the cosmology of the dark sector and the resulting constraints on our framework.  We find constraints on the size of the kinetic mixing between the dark sector and SM, $\epsilon$, on the DM interactions, and on the spectrum of the GeV scale states.  The cosmology of our model resembles the cosmology of the annihilating DM framework of Ref.~\cite{ArkaniHamed:2008qn}.  
For related discussions of the cosmology of GeV scale hidden sectors, see Refs.~\cite{Finkbeiner:2007kk,ArkaniHamed:2008qp,Cheung:2009qd,Finkbeiner:2009mi,Morrissey:2009ur}.
Below we include several new observations and a novel emphasis on the
aspects of the cosmology that are important for decaying DM\@.  We
begin this section by discussing the relic density of DM, and end with
a discussion on the cosmology of light dark sector fermions, which can
decay to observable gamma rays providing a smoking gun signature of decaying DM~\cite{Ruderman:2009ta}.

\subsection{Thermal DM Abundance}

A model of DM must of course reproduce the observed relic density, $\Omega_\chi h^2 \simeq 0.1$.  The `WIMP Miracle' implies that the correct abundance is achieved if DM is in kinetic equilibrium with the SM when it freezes out, with a WIMP cross-section, $\left< \sigma_\chi v \right> \simeq 3 \times 10^{-26} \unit{cm}^3 \unit{s}^{-1}$.  The same cosmology applies for decaying DM, as mentioned in the introduction, since the decay rate is much longer than the age of the Universe.  We now discuss how our model can satisfy these requirements.

DM retains the usual thermal history by interacting with dark gauge
bosons which are in kinetic equilibrium with the SM~\cite{Finkbeiner:2007kk,ArkaniHamed:2008qp,Finkbeiner:2009mi}.  The
kinetic equilibrium is maintained by interacting with the SM thermal
bath through the kinetic mixing, $\dgauge \, \psi_{\rm SM}
\leftrightarrow \gamma \, \psi_{\rm SM}$, where $\psi_{\rm SM}$
denotes any relativistic SM particle with hypercharge.  This reaction
remains efficient for temperatures in the range $\mdark \lesssim T_{\rm kin} \lesssim (\epsilon^2 \alphaEM^2 / \pi^2 g_*^{1/2}) \Mp$, where $g_*$ is the number of relativistic degrees of
freedom at temperature $T_{\rm kin}$.  For the DM to be a thermal relic with a
WIMP cross-section, $T_{\rm kin}$ must be larger than the DM
decoupling temperature, $T_{\rm dec} \simeq \mdm/20$.  The thermal history therefore places a lower-bound on the size of the kinetic mixing:
\begin{equation}
  \label{eq:epsilonconstraint}
  \epsilon \gtrsim 10^{-5} - 10^{-6}. 
\end{equation}
There is tension between this constraint, and the constraint on
$\epsilon$ from direct detection when DM couples elastically to the
dark photon, Eq.~\eqref{eq:directdetect}.  One way to evade the
constraint of Eq.~\eqref{eq:epsilonconstraint} is to introduce weak
scale particles charged under both the dark sector and the SM\@.
Particles charged under both sectors can maintain kinetic equilibrium,
but they must be very light, $\cO(100\unit{GeV})$,  in order to do so until $T_{\rm dec}$.  Another way to alleviate this tension is to introduce a DM splitting, which evades the constraint from direct detection.

The introduction of a DM splitting can change the DM annihilation
cross-section in an interesting way.  A splitting can be generated
radiatively or through Yukawa interactions, as we discuss in section
\ref{sec:mev-scale}.  Radiative splittings are generated after
breaking non-Abelian dark sectors with specific matter content,
however such models are significantly 
 more complicated to construct~\cite{Baumgart:2009tn}.  A simpler
 alternative, when the DM is charged under $\Udark$, is to couple it
 directly to the light Higgses, as in Eq.~\eqref{eq:split}.  In
 addition to introducing splittings, these interactions provide the
 DM a direct annihilation channel into light Higgses.  This Yukawa
 annihilation rate, $\sigma_y$, can be parametrically related to the
 annihilation rate into dark gauge bosons, $\sigma_g$, as
\begin{eqnarray}
  \label{eq:crosscompare}
  \frac{\sigma_y}{\sigma_g} \simeq 
  \left(\frac{\mdm}{\mdark}\right)^4\left(\frac{\delta
      \mdm}{\mdm}\right)^2 =
  \left(\frac{0.5\unit{GeV}}{\mdark}\right)^4\left(\frac{\mdm}{2.5\unit{TeV}}\right)^2\left(\frac{\delta
    \mdm}{100\unit{keV}}\right)^2. 
\end{eqnarray}
Here $\delta \mdm$ is the size of the DM splitting,
Eq.~\eqref{eq:splitsize}.  We see that the Yukawa annihilation channel
parametrically dominates the DM relic density when $\mdark \lesssim
500$ MeV or when the splitting is sufficiently large.  In this regime,
the DM gauge coupling must be small in order for the DM to have the
correct relic density.  This implies that non-perturbative Sommerfeld
enhancements to the annihilation cross-section are $\lesssim
\cO(100)$.  Decaying DM models in this regime evade the constraints
from the Sommerfeld enhancements discussed in section \ref{sec:dangers}, and annihilating models of this type cannot achieve a large enough Sommerfeld enhancement to fit FERMI~\cite{Meade:2009iu}.

A similar analysis applies when DM is charged under the SM and couples
to the $Z$, as in the model of section \ref{sec:SU5}.  A splitting is
required to evade the constraints from direct detection, which can be
introduced by coupling the DM to the SM Higgs.  This opens up a new annihilation channel of DM into SM Higgses, which allows for a larger annihilation cross-section and heavier DM masses, as discussed in section \ref{sec:SU5}.

An interesting example that has no tension between the thermal history
and direct detection, and does not require a DM splitting, is our
$\Utev \times \Udark$ model of section \ref{sec:u1u1}.  Here, the DM
is charged under $\Utev$, which is broken at the weak scale and
kinetically mixes with the GeV-scale dark sector, $\Udark$, with
mixing of order  $\epsilon_d \sim 10^{-3}$.  There is no strong
constraint on this model from direct detection because DM does not
couple directly to the $Z$ or the light dark photon.  The light
dark sector, $\Udark$, stays in kinetic equilibrium with the SM
through kinetic mixing, as discussed above.  The kinetic mixing
between $\Utev$ and $\Udark$ keeps $\Utev$, and therefore the  DM, in
kinetic equilibrium with $\Udark$ through the interaction $\gamma_\chi
\, h \leftrightarrow \dgauge \, h$, with $h$ corresponding to any of
the light Higgses or Higgsinos charged under the $\Udark$.  The DM is thus kept in kinetic equilibrium with the SM through double kinetic mixing, yielding the correct relic abundance\footnote{For this model, the DM does kinetically decouples from the SM  during freezeout at $T_{\rm kin} =m_{\gamma_\chi} \lesssim \mdm$.  After decoupling, the DM temperature scales as $T = T_\gamma^2 / T_{\rm kin}$, but this only modifies the relic density by an $\cO(1)$ amount.}.

\subsection{The Lightest Dark Sector Fermion}

The dark sector may contain light particles that are long-lived.  Such
fields are constrained by cosmology, as we discuss now.  There is
typically no constraint on light scalars and gauge bosons since both
can decay through the kinetic mixing with  cosmologically fast
timescales, as in Eq.~\eqref{eq:bosondecays}.  An exception to this
are stable light scalars due to an unbroken discrete symmetry, which
we discuss below.  The lightest fermion, on the other hand, must decay
to the gravitino, if kinematically allowed, which can lead to
cosmologically long lifetimes.  In what follows, we focus on the
situation where the lightest fermion mixes with the gaugino, and we
consider separately the cases where it is heaver than, approximately
degenerate with, or lighter than the dark gauge boson.  We show that
the last two scenarios require the lightest fermion to decay to a
photon and a gravitino on sub-galactic length scales, leading to observable gamma ray signatures~\cite{Ruderman:2009ta}.

\begin{itemize}

\item {$\mgaugino >  \mdark$}

This regime applies when there is sizeable SUSY breaking in the dark
sector $\gtrsim $ GeV\@.  The fermions can annihilate into dark gauge
boson pairs, with cross-section $\sigma \simeq \gdark^4/(8\pi
\textrm{GeV}^2)$ which leads to an abundance $\Omega_{\dgaugino}
h^2\simeq 10^{-6}$.  After freezeout, the fermion can decay to the
dark gauge boson and a gravitino, which is kinematically allowed for
low scale SUSY breaking, $\sqrt F \lesssim 10^9$ GeV\@.  The dark gauge
boson then decays through the kinetic mixing to leptons.   The
corresponding fermion lifetime is given by
Eq.~\eqref{eq:decaytodarkphoton}.   For an abundance this small, there
is no constraint from BBN for electromagnetic
decays~\cite{Jedamzik:2006xz}.  There are, on the other hand, strong
constraints on electromagnetic decays after
recombination~\cite{Kribs:1996ac}, however the decay discussed above
always proceeds before recombination and hence evades the bound.  An analogous discussion applies if the lightest dark fermion is a Higgsino that is heavier than its scalar superpartner.  We conclude that the dark sector is not constrained by the lightest fermion when it is heavier than its superpartner.

\item {$\mgaugino \sim \mdark$}

Let us now consider  the regime where the dark gaugino is
approximately degenerate with the dark gauge boson.  This is the case
when the dark sector spectrum is approximately supersymmetric, for
instance when D-term mixing dominates as discussed in
section~\ref{sec:gev-scale}.  When the temperature is above $\mdark$,
the dark sector is in kinetic equilibrium with the SM and the number
density of the dark bosons and fermions are of  the same order of
magnitude.  When the temperature drops below $\mdark$, the dark gauge
bosons cannot be created from the thermal bath, and they decay
instantly to SM leptons through the kinetic mixing, on a timescale
much faster than the Hubble rate, as in Eq.~\eqref{eq:bosondecays}.
The dark gauginos, on the other hand, are long lived with abundance
controlled by their available annihilation channels.  As long as
$\mgaugino > \cO(0.85) \, \mdark$, the finite temperature allows dark
gauginos to annihilate into a dark gauge boson pairs~\cite{Griest:1990kh}, with cross-section
 \begin{eqnarray}
 \label{eq:simgauginosigma}
  \langle\sigma_{\dgaugino}  v\rangle & \simeq & \cO(0.1)\times 
 \frac{\gdark^4}{8 \pi \mgaugino^2} \simeq 10^{4} \
 \langle\sigma_\chi v \rangle \left( \frac{\gdark}{0.35} \right)^4 \left(\frac{1 \unit{GeV}}{\mgaugino} \right)^2,
\end{eqnarray}
where the $\cO(0.1)$ suppression results from thermal averaging, and $\sigma_\chi$ is the DM annihilation cross-section. 

For the above parameters, the resulting relic density is
$\Omega_{\dgaugino} h^2 \simeq 10^{-5}$.  The gauginos will decay to
photons and gravitinos with lifetime given by
Eq.~\eqref{eq:decaytophoton}.  For an abundance of this size, there is
no constraint from BBN on the resulting electromagnetic decays (see
Fig. 9 from the first reference of~\cite{Jedamzik:2006xz}), but the gauginos must decay before
recombination, $\tau < 10^{13}$ sec, to avoid constraints from diffuse
gammas~\cite{Kribs:1996ac}.  Amusingly, there is a coincidence in
which the time of recombination roughly equals the amount of time it
takes light to cross our Galaxy.  As a consequence, the constraint
from recombination guarantees that dark gauginos produced in our
galaxy decay to observable gamma rays.  The resulting constraint on
the size of the kinetic mixing is $\epsilon \gtrsim 10^{-9}$ and fixing $\epsilon \simeq 5 \times 10^{-4}$, the constraint on the SUSY breaking scale is $\sqrt F \lesssim 2\times10^7$ GeV\@.  A possible caveat in the above argument, is  that by the time the fermions decouple, the gauge boson are already kinetically decoupled 
from the thermal bath. This may alter the final abundance by some (order one) amount. A 
better understanding requires solving the exact Boltzmann equations, which is beyond the 
scope of this paper.

\item {$\mgaugino < \cO (0.85) \, \mdark$}

  Lastly, we consider the regime where the dark gaugino is
  significantly lighter than the dark gauge boson, which as in the
  first case requires
  GeV-scale SUSY breaking in the dark sector.  If $\mgaugino \gtrsim
  0.5 \, \mdark$, a gaugino pair can annihilate into one dark gauge
  boson, and an $e^+ e^-$ pair, through kinetic mixing.  The resulting
  cross-section is suppressed by $\epsilon^2$,
\begin{eqnarray}
  \label{eq:lightestgauginosigma}
 \left< \sigma_{\dgaugino} v \right> \simeq
  \epsilon^2 \alphaEM \frac{\gdark^4}{8 \pi \mgaugino^2} &=& 10^{-4} \left< \sigma_\chi v \right> \left( \frac{\epsilon}{5 \times 10^{-4}} \right)^2  \left( \frac{\gdark}{0.35} \right)^4 \left(\frac{1 \unit{GeV}}{\mgaugino} \right)^2.
\end{eqnarray}
The abundance is $\Omega_{\dgaugino} h^2 \simeq 10^3$, and the BBN
constraint now requires $\tau_{\dgaugino} \lesssim 10^4$ sec.  This
constraint is rather strong and can be marginally satisfied for the
parameters of Eq.~\eqref{eq:decaytophoton}.  We see that rather large kinetic mixing and a low SUSY breaking scale are both necessary.  Again, the gaugino decays to an observable gamma ray.  Finally, we note that when $\mgaugino < \mdark / 2$, the gauginos must annihilate into $2e^+2 e^-$, with cross-section suppressed by an additional $\epsilon^2 \alphaEM$ relative to Eq.~\eqref{eq:lightestgauginosigma}, ruling out models where the lightest fermion is lighter than $\mdark/2$.

\end{itemize}

To summarize our findings, we see that our model is unconstrained by the lightest fermion when $\mgaugino \gtrsim \mdark$, and that otherwise cosmological constraints imply that the dark gaugino must decay to gamma rays with short lifetimes compared to galactic length scales, leading to observable gamma ray signatures.  These constraints are manifest as limits on the size of kinetic mixing, $\epsilon$, and the SUSY breaking breaking scale $\sqrt F$, as we discuss above.  

We conclude this section by noting that there may be light particles
that are completely stable.  For example, in the model of section
\ref{sec:U1}, $h$ and $n$ are stable, which follows from their charges
under an unbroken $\Z_2$.  If $h$ and $n$ are heavier than the dark
gauge boson, they have a large annihilation cross-section which is
parametrically similar to the heavy gaugino case discussed above, thus
resulting in a small relic density, $\Omega_h \simeq 10^{-6}$.  On the
other hand, if $h$ and $n$ are lighter than the dark gauge boson, they
will have a large abundance and the model is excluded.  In general,
light fields that are stable due to discrete symmetries must be
heavier than, and annihilate into, the unstable and lighter dark sector fields.

\section{Discussion}
\label{sec:conclusions}

The decaying DM models proposed in this paper predict a number of
signals at upcoming experiments.  The light dark sector particles can
be produced in colliders, resulting in lepton jets, as in the
annihilating models
of~\cite{ArkaniHamed:2008qn,ArkaniHamed:2008qp,Baumgart:2009tn}.  The
dark sector can also be probed at low energy $e^+ e^-$ colliders and
fixed target experiments~\cite{Pospelov:2008zw}.  These direct
production experiments have the potential to discover the dark sector,
but probably cannot tell apart decaying and annihilating models.  On
the other hand, astrophysical signals can differentiate between the
two scenarios and provide a complementary means to probe the dark
spectrum~\cite{Ruderman:2009ta}.  As we discuss above, primary photons
are produced when the dark gaugino is degenerate with or lighter than
the dark photon.  This results in a hard gamma ray spectrum that can
be discovered by HESS, AGIS, and CTA and possibly FERMI~\cite{Ruderman:2009ta}.  Moreover, if
DM is charged under the SM, as in the model of section \ref{sec:SU5},
decays produce primary neutrinos, resulting in a hard neutrino
spectrum that can be measured at upcoming experiments such as
IceCube/DeepCore.  The situation is distinct from the annihilating
models.  For those, measurements from the GC exclude the production of
primary photons and neutrinos with sizeable branching
fractions~\cite{Meade:2009iu,Bergstrom:2009fa,Meade:2009rb,Bergstrom:2008ag}.

We conclude with two further directions that can be explored in these models.
\begin{itemize}
\item It would be interesting to construct a model that is more
  directly related to the SUSY breaking sector.  We have taken DM to
  receive a weak scale mass by coupling it to a singlet.  Since the DM
  is not required to be charged under the SM or under the dark sector,
  another interesting possibility is for the DM to reside in the SUSY
  breaking sector, for example as a pseudomodulus~\cite{Shih:2009he}.
\item The $\Udark$ and $SU(2)_d$ models in sections~\ref{sec:U1}
  and~\ref{sec:SU2} respectively, include two species of DM $\chi_1$
  and $\chi_2$.  The existence of several species has several
  interesting implications.  First, there can be
  `Wimponium'~\cite{MarchRussell:2008tu} bound states, $\chi_1 \bar
  \chi_2$ and $\chi_2 \bar \chi_1$, which are cosmologically
  long-lived.  Second, it may be possible to include both the iDM and
  XDM proposals since we have shown that both species can have
  MeV-sized DM splittings.  The viability of these ideas requires
  further study.
\end{itemize}

\ack
We thank N.~Arkani-Hamed, C.~Cheung, P.~Meade, D.~Morrissey,
M.~Papucci, T.~Slatyer, N.~Weiner, and I.~Yavin for useful
discussions.  We thank D.~Shih for discussions and comments on the manuscript.  
 J.~T.~R. is supported by an NSF graduate fellowship.  T.~V. is supported by DOE grant DE-FG02-90ER40542.

\appendix

\renewcommand{\theequation}{A-\arabic{equation}}
\setcounter{equation}{0}

\section{The Models: Superpotentials and Charges}
\label{app:ModelCharges}
In each of the models of section \ref{sec:Models}, the DM can decay
through dimension-6 GUT suppressed operators into GeV states in the
hidden sector.  For these models to work, it is necessary that renormalizable or dimension-5
operators that allow DM decays are absent.  Such dangerous decays can be forbidden
by global discrete symmetries at the GUT scale.
Such symmetries also forbid a GUT scale mass for the 
DM\@.  In this appendix we verify that the above models are generic
and safe, by presenting such  global symmetries that forbid both
dangerous DM decays and GUT scale masses for light fields.  We also
collect the full superpotentials of each model, for easy reference.

\subsection*{$\Udark$}

The superpotential of our minimal $\Udark$ model is given by:
\begin{eqnarray}
W & = & W_{\rm decay} + W_{\rm DM} + W_{\rm split}\,, \nonumber \\
W_{\rm decay} &=& \left(M_{\rm GUT} + X\right) Y \bar Y\nonumber +  M_{\rm GUT} X \bar X + \bar X \chi_1 \bar \chi_2 \,, \nonumber \\
W_{\rm DM} & = &S\left( \chi_1 \bar \chi_1+ \chi_2 \bar \chi_2 \right) + n h \bar h \,, \nonumber \\
W_{\rm split} & = & \sum_{i=1}^2 \left(S N_i^2+N_i \chi_i \bar h \right)\,.
\end{eqnarray}
There are several dangerous operators that are allowed by $\Udark$
gauge invariance.  These include a GUT scale mass of the form, $\chi_i \bar
\chi_j$, a TeV scale mass for the light Higgses, $S h \bar h$,
renormalizable DM decay, $\chi_2 \bar h n$, and dimension-5 DM decay operators,
$\chi_2 h \bar h^2$.  All dangerous operators of these types are
forbidden by the $\Z_4^R \times \Z_4$ symmetry displayed in the upper
left of table~\ref{tab:charge}.

\subsection*{$\SUtwodark \rightarrow \Udark$}
\label{app:SU2}

The superpotential of our $SU(2)_d \rightarrow \Udark$ model is:
\begin{eqnarray}
W &= & W_{\rm decay} + W_{\rm GUT} + W_{\rm DM} + W_{\rm split}\,, \nonumber \\
W_{\rm decay} &=& f(H) + H X^2 \,, \nonumber \\
W_{\rm GUT} &=& \tr \left[g(H) \left( \Phi\bar\Phi+ S_\Phi \bar\Phi + n^\prime\bar n +  s_n\bar n^\prime + s_N N + s_{\bar N} \bar N \right) \right]\,, \nonumber \\
W_{\rm DM} &=& (\Phi + S_\Phi) \chi \bar \chi  + (n^\prime+s_n) h^2 \,, \nonumber \\
W_{\rm split} & = & (\Phi + S_\Phi) N \bar N + N (\chi^2 + \bar \chi^2) + \bar N h^2\,.
\end{eqnarray}
$W_{\rm split}$ is not discussed in section~\ref{sec:SU2}, and is
necessary to generate a DM splitting that evades the constraints from
direct detection, as described in section~\ref{sec:dangers}.  We also
add the final two terms to $W_{\rm GUT}$.  Once $H$ obtains a VEV, $N_3$
and $\bar N_3$ receive GUT scale masses while the charged components
remain light.  At low energies, $W_{\rm split}$ takes the form
\begin{equation}
W_{\rm split}^{\rm eff} = S (N_- \bar N_+ + N_+ \bar N_-) + N_-(\chi_1^2+ \bar \chi_2^2) + N_+ (\chi_2^2 + \bar \chi_1^2)+\bar N_- h_1^2 + \bar N_+ h_2^2,
\end{equation}
where $S$ is the light linear combination of $\Phi_3$ and $S_\Phi$, as
in Eq.~\eqref{eq:SU2Wdmeff}.  Expanding around the true minimum with
$\langle S\rangle \sim \unit{TeV}$ and $\langle h_2\rangle \sim
\unit{GeV}$, we see that $\bar N_-$ has a tadpole term which induces a
VEV for $N_-$ of order$\unit{GeV}^2/\unit{TeV}$.  Consequently, $\langle
N_-\rangle$ contributes to the mass of $\chi_1$ and $\bar\chi_2$,
which splits the $\chi_i$ and $\bar\chi_i$ multiplets.  These splittings allow the model to evade
the constraints from direct detection, and to possibly incorporate the
iDM and/or XDM proposals.  In the upper right of
table~\ref{tab:charge}, we display a $\Z_{16}$ symmetry which forbids
GUT scale masses for light fields and dangerous decays for DM\@.

 In order to avoid a Landau pole below the GUT scale, the field content
of this model requires that $\alphadark \lesssim 1/100$.  If DM
annihilates only into light gauge fields, a gauge coupling of this
size is insufficient to produce the correct DM relic density.
Fortunately, $W_{\rm split}$ introduces DM annihilations into the
light Higgses, which can dominate the annihilation cross-section and
lead to the correct relic density.

\subsection*{SM Charged DM: $\SUfive \times \Udark$}

The superpotential and \kahler term of our model
with DM charged under the SM are given by:
\begin{eqnarray}
W & = & W_{\rm decay} + W_{\rm DM} + W_{\rm SM}\,, \nonumber \\
W_{\rm decay} &=&  (\mgut + X) Y \bar Y + \mgut X \bar X + \bar X \chi{\bf \bar 5}_f \,,\nonumber \\
W_{\rm DM} & = & S \left( \chi \bar \chi +N^2 + s_1^2 \right) + \chi H_d N + n h \bar h \,,\nonumber \\
W_{\rm SM} & = & S H_u H_d + {\bf 10}_f {\bf \bar 5}_f H_d + {\bf 10}_f^2 H_u + \frac{H_u^2 {\bf \bar 5}_f^2} {M_{\rm GUT}} \,,\nonumber \\
K & \supset & \frac{\bar \chi {\bf \bar 5}_f^\dagger s_1} {M_{\rm GUT}}\,.
\end{eqnarray}
Here $W_{\rm SM}$ denotes the usual $SU(5)$ GUT superpotential with
Majorana neutrino masses and the NMSSM singlet for generating the $\mu$ term.
Dangerous decay operators now include renormalizable Yukawa couplings
between DM and the SM, such as $10_f \bar \chi H_d$.  Such operators
must be forbidden, and for this reason DM cannot be a fourth flavor.
In table \ref{tab:charge}, we list the charges under a $\Z_{2}^R
\times \Z_{3}^R \times Z_{6}$ symmetry that forbids all dangerous
decays and GUT scale masses, where the $\Z_{2}^R$ extends the usual
R-parity to the new fields.

\subsection*{$\Utev \times \Udark$} 

The superpotential of our $\Utev \times \Udark$ model is given by: 
\begin{eqnarray}
W & = & W_{\rm decay} + W_{\rm DM}\,,  \nonumber \\
W_{\rm decay} &=& \left(M_{\rm GUT} + X\right) Y \bar Y\nonumber +  M_{\rm GUT} X \bar X + \bar X \chi_1 \bar \chi_2 \,, \nonumber \\
W_{\rm DM} & = & S_2 \, \chi_2 \bar \chi_2  +S_1(\chi_1 \bar \chi_1 + S_2^2) + n h \bar h\,. 
\end{eqnarray}
This model is particularly simple since no DM splitting is required to
evade the constraints from direct detection.  There is a $\Z_{9}^R$
symmetry, listed in the lower left side of table~\ref{tab:charge}, that
forbids both dangerous DM decays and GUT scale masses for the light
fields.

\begin{table}[hp!]
\vspace{-1cm}
\footnotesize
\hbox{\hspace{3cm}
\begin{tabular}{|c c||c| c c| }
\hline
&& $\Udark$ & $\Z_4^R$ & $\Z_4$ \\
\hline \hline
\multirow{4}{*}{\begin{sideways}GUT\end{sideways}}
& $X$ & 0&0&0\\
&$\bar X$&0&2&0\\
& $Y$ & 1 & 0 & 0 \\
& $\bar Y$ &-1& 2& 0 \\
\hline
\multirow{7}{*}{\begin{sideways}TeV\end{sideways}}&$\chi_1$ & 1&2&3\\
&$\bar \chi_1$ & -1 & 0 &3 \\
&$\chi_2$ &1&0&1\\
&$\bar \chi_2$ &-1 &2&1\\
&$S$ & 0 &0&2\\
&$N_1$&0&3&3\\
&$N_2$&0&1&1\\
\hline
\multirow{3}{*}{\begin{sideways}GeV\end{sideways}}&$h$ & 1&1&1\\
&$\bar h$&-1&1&2\\
&$n$&0&0&1\\
\hline
\end{tabular}

\hspace{2cm}

\begin{tabular}{|c c||c| c | }
\hline
&& $SU(2)_d$ & $\Z_{16}$  \\
\hline \hline
\multirow{4}{*}{\begin{sideways}GUT\end{sideways}}
& $H$ & {\bf Adj} & 0  \\
& $X$ & $\fund$ & 8 \\
& $ \bar \Phi$ & {\bf Adj} &2\\
&$\bar n$&{\bf Adj}&4\\
\hline
\multirow{6}{*}{\begin{sideways}TeV\end{sideways}}
&$\chi$ & $\fund$&13\\
&$\bar \chi$ & $\fund$& 5  \\
&$\Phi$ &{\bf Adj}& 14\\
&$S_\Phi$ & {\bf 1} &14\\
&$N$&{\bf Adj} &6\\
&$\bar N$&{\bf Adj}&12\\
\hline
\multirow{3}{*}{\begin{sideways}GeV\end{sideways}}
&$h$ & $\fund$&2\\
&$n^\prime$&{\bf Adj}&12\\
&$s_n$&{\bf 1}&12\\
\hline
\end{tabular}}

\vspace{1 cm}

\hbox{\hspace{2cm}
\begin{tabular}{|c c||c c|  c| }
\hline
&& $\Utev$ & $\Udark$ & $\Z_{9}^R$ \\
\hline \hline
\multirow{4}{*}{\begin{sideways}GUT\end{sideways}}
& $X$ & 0&0&0\\
&$\bar X$&0&0&2\\
& $Y$ & 0 & 1 & 8 \\
& $\bar Y$ &0& -1& 3 \\
\hline
\multirow{6}{*}{\begin{sideways}TeV\end{sideways}}
&$\chi_1$ & 1&0&0\\
&$\bar \chi_1$ & -1 & 0 &5 \\
&$\chi_2$ & 1&0&4\\
&$\bar \chi_2$ & -1 & 0 &0 \\
&$S_1$ &0&0&6\\
&$S_2$ &0&0&7\\
\hline
\multirow{3}{*}{\begin{sideways}GeV\end{sideways}}
&$h$ & 0&1&2\\
&$\bar h$&0&-1&6\\
&$n$&0&0&3\\
\hline
\end{tabular}

\hspace{2cm}

\begin{tabular}{|c c||c c|  c c c| }
\hline
&& $SU(5)$ & $\Udark$ & $\Z_{2}^R$ & $\Z_{3}^R$ & $\Z_{6}$ \\
\hline \hline
\multirow{4}{*}{\begin{sideways}GUT\end{sideways}}
& $X$ & {\bf 1}&0&0&0&0\\
&$\bar X$&{\bf 1}&0&0&2&0\\
& $Y$ & {\bf 1} & 1 & 0&2&2 \\
& $\bar Y$ &{\bf 1}& -1& 0&0&4 \\
\hline
\multirow{5}{*}{\begin{sideways}TeV\end{sideways}}
&$\chi$ & $\fund$&0&1&2&4\\
&$\bar \chi$ & $\overline \fund$ & 0 &1&1&0 \\
&$S$ &{\bf 1}&0&0&2&2\\
&$N$&{\bf 1}&0&1&0&2\\
&$s_1$ &{\bf 1}&0&0&0&2\\
\hline
\multirow{3}{*}{\begin{sideways}GeV\end{sideways}}
&$h$&{\bf 1}&1&1&0&4\\
&$\bar h$&{\bf 1}&-1&0&1&1\\
&$n$&{\bf 1}&0&1&1&1\\
\hline
\multirow{4}{*}{\begin{sideways}SM\end{sideways}}
&$H_u$&$\fund$&0&0&0&4\\
&$H_d$&$\overline \fund$&0&0&0&0\\
&${\bf \bar 5}_f$&$\overline \fund$&0&1&1&2\\
&${\bf 10}_f$&\begin{tabular}{c} $\asymm$ \end{tabular}&0&1&1&4\\
\hline
\end{tabular}}
\caption{The gauge charges and example global charges for each model.  Clockwise from the upper left, are the $\Udark$, $SU(2)_d \rightarrow U(1)_d$, $SU(5)_{\rm SM} \times \Udark$, and $\Utev \times \Udark$ models.  For each model, the charges forbid renormalizable and dimension 5 dark matter decays and GUT scale masses for light fields.}
\label{tab:charge}
\end{table}

\newpage

 \footnotesize

\begin{multicols}{2}

\end{multicols}

\end{document}